\documentclass[sigconf,9pt,numbers,nonacm]{acmart} 
\usepackage{lipsum}
\usepackage{enumitem}
\usepackage{subfigure}
\usepackage{blindtext}

\usepackage{siunitx}
\usepackage{subcaption}
\usepackage{array}
\usepackage{algorithm}
\usepackage{algpseudocode}
\usepackage{makecell}
\usepackage{pifont}
\usepackage{tikz}
\usepackage{xcolor}
\usepackage{threeparttable}
\usepackage{booktabs}
\usepackage{colortbl}
\usepackage{array,adjustbox,tabularx}
\newcolumntype{Y}{>{\centering\arraybackslash}X}
\renewcommand\footnotetextcopyrightpermission[1]{} 
\definecolor{myred}{RGB}{230,30,28}
\definecolor{myyellow}{RGB}{244,220,108}
\newcommand{\cmark}{\textcolor{green!60!black}{\ding{51}}}  
\newcommand{\xmark}{\textcolor{red}{\ding{55}}}             
\newcommand{\sssec}[1]{\vspace*{0.05in}\noindent\textbf{#1}}
\newcommand{\sysname}{\textit{GigaFlex}}
\newcolumntype{C}{>{\centering\arraybackslash}X}
\newcolumntype{L}{>{\raggedright\arraybackslash}X} 

\AtBeginDocument{%
  }

\setcopyright{acmlicensed}
\copyrightyear{2026}
\acmYear{2026}
\acmSubmissionID{13}
\acmDOI{XXXXXXX.XXXXXXX}

\newcommand{\squishlist}
{\begin{itemize}[itemsep=1pt,parsep=2pt,topsep=3pt,partopsep=0pt,leftmargin=0em, itemindent=1em,labelwidth=1em,labelsep=0.5em]}
\newcommand{\squishend}{\end{itemize}}
\newcommand{\squishenum}{%
  \begin{enumerate}[itemsep=1pt,parsep=2pt,topsep=3pt,partopsep=0pt,
  leftmargin=0em,itemindent=1.5em,labelwidth=1em,labelsep=0.5em,
  label=\textbf{(\arabic*)}]}

\newcommand{\squishsubenum}{%
  \begin{enumerate}[itemsep=1pt,parsep=2pt,topsep=0pt,partopsep=0pt,
  leftmargin=0em,listparindent=1.5em,labelwidth=1em,labelsep=0.5em,
  label=\textbf{(\alph*)}]}

\newcommand{\squishenumend}{\end{enumerate}}

\usepackage{graphicx}
\newcommand{\chaos}{\includegraphics[height=1em]{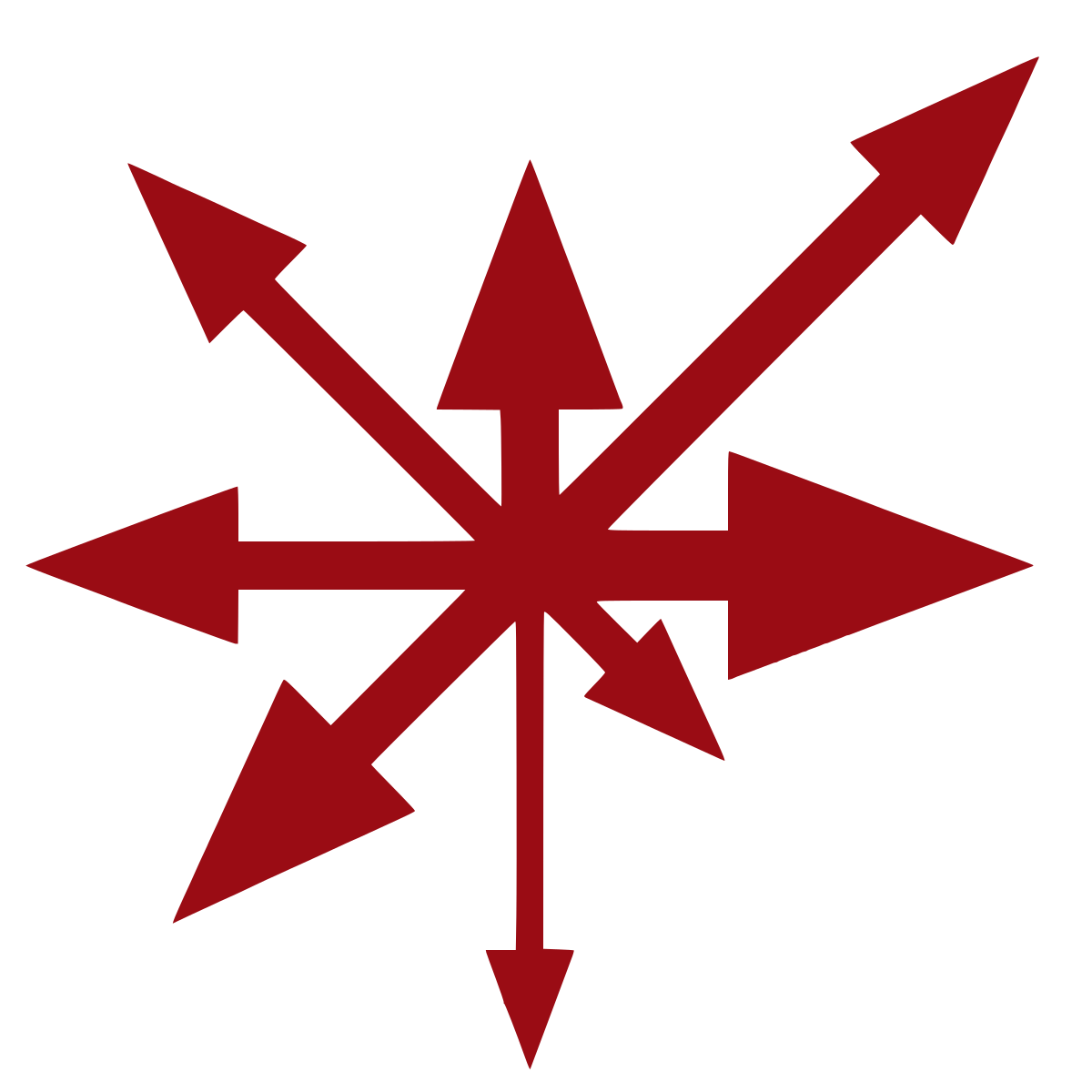}}

\author{Jiangyifei Zhu$^{\dag}$}
\affiliation{  
  \institution{Carnegie Mellon University}
  \city{Pittsburgh}
  \state{PA}
  \country{USA}}
\email{jiangyiz@andrew.cmu.edu}

\author{Yuzhe Wang$^{\dag}$}
\affiliation{  
  \institution{Carnegie Mellon University}
  \city{Pittsburgh}
  \state{PA}
  \country{USA}}
\email{yuzhew@andrew.cmu.edu}

\author{Tao Qiang$^{\dag}$}
\affiliation{  
  \institution{Shanghai Jiao Tong University}
  \city{Shanghai}
  \country{China}}
\email{riderdecade@sjtu.edu.cn}

\author{Vu Phan}
\affiliation{  
  \institution{Carnegie Mellon University}
  \city{Pittsburgh}
  \state{PA}
  \country{USA}}
\email{vuphan@andrew.cmu.edu}

\author{Zhixiong Li}
\affiliation{  
  \institution{Carnegie Mellon University}
  \city{Pittsburgh}
  \state{PA}
  \country{USA}}
\email{zhixionl@andrew.cmu.edu}

\author{Evy Meinders}
\affiliation{  
  \institution{Carnegie Mellon University}
  \city{Pittsburgh}
  \state{PA}
  \country{USA}}
\email{emeinder@andrew.cmu.edu}

\author{Eni Halilaj}
\affiliation{  
  \institution{Carnegie Mellon University}
  \city{Pittsburgh}
  \state{PA}
  \country{USA}}
\email{ehalilaj@andrew.cmu.edu}

\author{Justin Chan}
\affiliation{  
  \institution{Carnegie Mellon University}
  \city{Pittsburgh}
  \state{PA}
  \country{USA}}
\email{justinchan@cmu.edu}

\author{Swarun Kumar}
\affiliation{  
  \institution{Carnegie Mellon University}
  \city{Pittsburgh}
  \state{PA}
  \country{USA}}
\email{swarun@cmu.edu}

\begin{document}

\title{{\sysname}: Contactless Muscle Exercise Monitoring using Radar}
\title{{\sysname}: Contactless Decoding of Muscle Vibrations\\Using a Chaos-Inspired Radar \chaos{}}
\title{{\sysname}: Contactless Monitoring of Muscle Vibrations\\During Exercise with a Chaos-Inspired Radar \chaos{}}

\begin{abstract}
In this paper, our goal is to enable quantitative feedback on muscle fatigue during exercise to optimize exercise effectiveness while minimizing injury risk. We seek to capture fatigue by monitoring surface vibrations that muscle exertion induces. Muscle vibrations are unique as they arise from the asynchronous firing of motor units, producing surface micro-displacements that are broadband, nonlinear, and seemingly stochastic. Accurately sensing these noise-like signals requires new algorithmic strategies that can uncover their underlying structure.  We present {\sysname} the first contactless  system that measures muscle vibrations using mmWave radar to infer muscle force and detect fatigue. \sysname\ draws on algorithmic foundations from Chaos theory to model the deterministic patterns of muscle vibrations and extend them to the radar domain. Specifically, we design a radar processing architecture that systematically infuses principles from Chaos theory and nonlinear dynamics throughout the sensing pipeline, spanning localization, segmentation, and learning, to estimate muscle forces during static and dynamic weight-bearing exercises. Across a 23-participant study, \sysname{} estimates maximum voluntary isometric contraction (MVIC) root mean square error (RMSE) of 5.9\%, and detects one to three Repetitions in Reserve (RIR), a key quantitative muscle fatigue metric, with an AUC of 0.83 to 0.86, performing comparably to a contact-based IMU baseline. Our system can enable timely feedback that can help prevent fatigue-induced injury, and opens new opportunities for physiological sensing of complex, non-periodic biosignals.

\begingroup
\renewcommand\thefootnote{}\footnotetext{$^\dag$~Equal-contribution first authors.}
\endgroup
\end{abstract}

\maketitle
\section{Introduction}

\renewcommand{\sssec}[1]{\vspace*{0.05in}\noindent\textbf{#1}}

\begin{center}
\textit{``If you get tired, learn to rest, not to quit.'' - Attributed to Banksy}
\end{center}

\begin{figure}[t]
    \centering
    \includegraphics[width=0.8\linewidth]{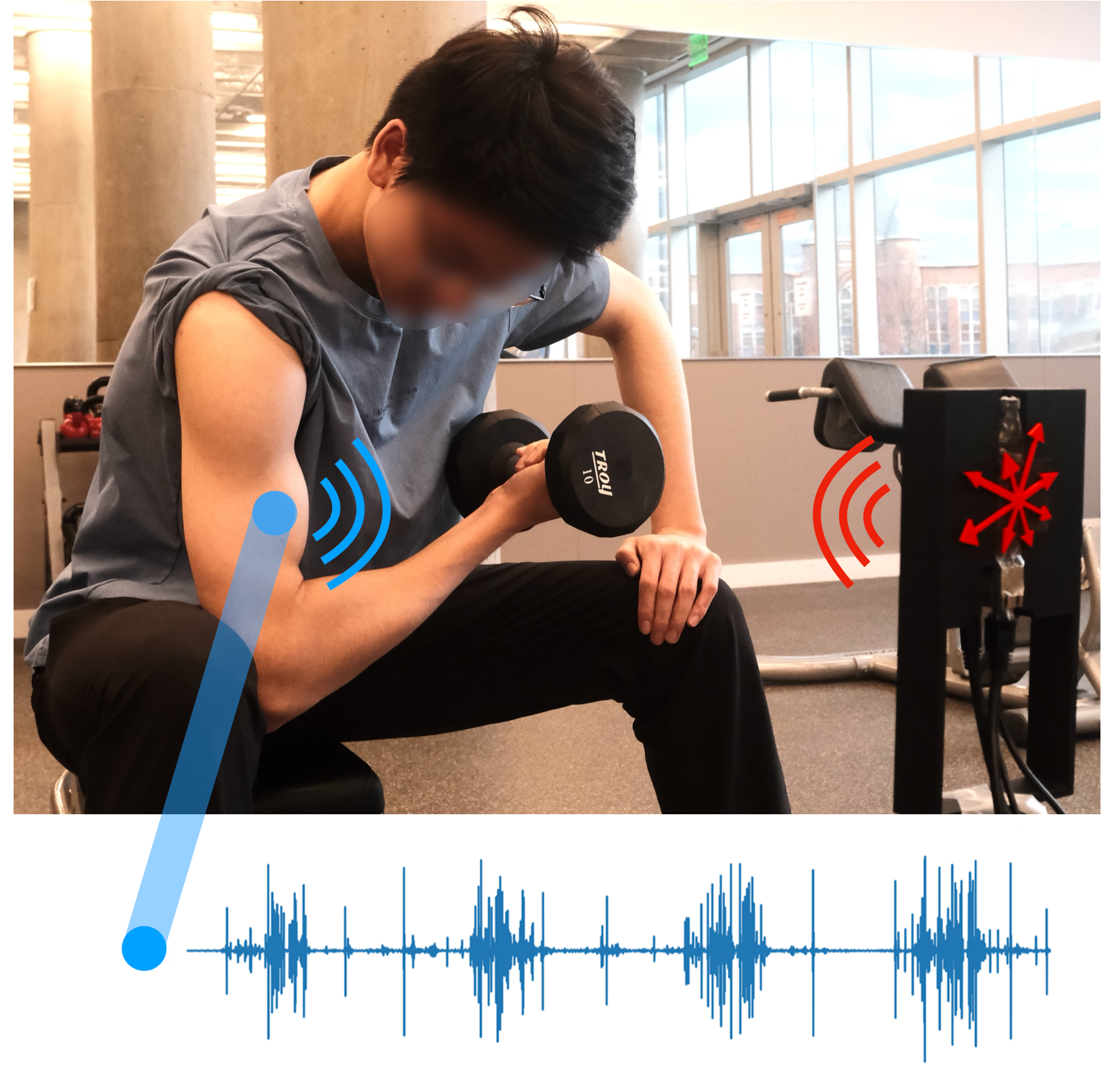}
    \vspace{-1em}
    \caption{{\sysname} continuously monitors muscle fatigue and force to help users maximize workout effectiveness while minimizing the risk of injuries with long recovery periods.}
    \label{fig:myovoid}
\end{figure}

In this paper we ask -- ``Can we sense and track muscle state during exercise without any contact-based sensors?'' Muscle state monitoring is key to a variety of contexts from exercise at the gym, rehabilitation and physical therapy. Our specific focus is on measuring muscle exertion levels and fatigue~\cite{lourenco_relationship_2023, Loy_Relatio_2017}. Consider for example, exercise at the gym where for an effective muscle workout one needs to be in the Goldilocks zone -- too little exertion of muscles renders the workout ineffective and too much may cause injury. Indeed, gym-related musculoskeletal injuries are common, with prevalence rates ranging about 35\% across multiple national studies~\cite{islam_prevalence_2024}. There is a significant proportion associated with over-lifting or misjudged muscle effort~\cite{lourenco_relationship_2023, Loy_Relatio_2017, islam_prevalence_2024} with recovery often taking several months~\cite{lubetzky2009prevalence,islam_prevalence_2024,sharma_prevalence_2024,alrushud_cross-sectional_2024}. Further, muscle state misperception worsens as exhaustion progresses or in users with recent injuries or muscle impairments~\cite{sayyadi_effectiveness_2024}. Hence, obtaining personalized and data-driven quantitative metrics for users to decide between ``taking a break'' versus ``lifting more'' is crucial to a safe and effective workout. 


Prior work on muscle state monitoring has largely relied on contact-based sensors. Many such sensors record electrical signals through surface electrodes~\cite{hermens2000development} and invasive needles~\cite{daube2009needle}. Others sense visible and microscopic changes to muscle geometry, such as bulging and vibrations are strongly correlated with electrical muscle levels and output ~\cite{hallock2021toward,shi2008continuous,kamatham2022simple,hallock2020muscle}. A variety of on-body sensors sense muscle vibrations and deformations using ultrasound imaging~\cite{shi2008continuous,brausch2022classifying,kamatham2022simple}, accelerometers~\cite{woodward2019segmenting}, and microphones~\cite{barry1985acoustic}.  However these approaches can be uncomfortable to wear for long periods and challenging to use in everyday environments. While less-common, contactless approaches to muscle monitoring have been explored including laser-doppler vibrometry~\cite{Sara_LDV} and active sonar~\cite{song_myomonitor_2021} but only focus on a single narrow point and require precise alignment. 


This paper presents \sysname{}, a mmWave radar-based system for contactlessly measuring muscle output at the arm without any body-worn devices or tags. We are inspired by rich recent work on physiological sensing using radar, for e.g.: for heartbeats~\cite{Langcheng_mmArrhythmia, fadel_smartHome}, pulse~\cite{geng_CaPTT_2023, liang_airbp_2023, zhu2025measuring}, and breathing~\cite{zhao2018noncontact,li2022detection}. However, sensing muscle vibrations and quantifying fatigue levels through radar is fundamentally more challenging, and to the best of our knowledge, a problem we are first to explore. Perhaps closest to our work is prior radar-based muscle activation sensing~\cite{tsengwa_toward_2024} which does not quantify muscle output, a key metric to decide when to rest or adjust effort to prevent injury~\cite{lubetzky2009prevalence,islam_prevalence_2024,sharma_prevalence_2024}. 

\sysname{}'s primarily technical challenge is that unlike breathing and heartbeats which are periodic and linear, muscle vibrations  are inherently aperiodic and nonlinear~\cite{gitter1995fractal,nieminen1996evidence}, arising due to asynchronous firing of motor units during muscle contraction~\cite{enoka_muscle_2008}. The resulting vibrations are bursty and broadband in nature, appearing stochastic at first glance (Fig.~\ref{fig:broadband}). To mitigate this, we design a radar sensing architecture that systematically infuses techniques from Chaos theory and nonlinear dynamics to reveal the hidden and deterministic structure of muscle vibrations. Our architecture builds on rich biomechanics literature~\cite{khodadadi2023nonlinear,xie2009detection,filligoi2011chaos,xiong2014application,conte2015chaos,lei2012nonlinear} where Chaos theory has successfully modeled muscle dynamics measured from accelerometers and electrodes. However, to the best of our knowledge, we are the first to adapt these techniques to the domain of radar sensing, and to extend them across our sensing pipeline. Through this architecture, we are able to leverage a \textit{single} radar device to produce continuous measures of muscle output for static isometric exercises (e.g. static elbow flexion), and dynamic isotonic exercises (e.g. concentric bicep curls).

\noindent The rest of this paper addresses the key technical components of our design: (1) Spatial Localization -- that tracks down radar signal components that spatially correspond to muscles; (2) Temporal Segmentation -- that extracts relevant chunks of the signal relevant to muscle activation, free of noise or other vibrations; (3) Muscle State Learning -- that processes the extracted signal to estimate muscle force levels and quantitative fatigue metrics. 

\sssec{(1) Chaos-Inspired Spatial Localization: } \textit{First, the inherent broadband and nonlinear nature of muscle vibrations makes it challenging to localize and distinguish them from noise, reflections, motion, and other biosignals, using conventional time-frequency filters.} To address this challenge, we leverage the inherent chaos in muscle vibrations to tell apart muscles from other forms of body movements and noise. Specifically, we empirically characterize the nonlinear properties of radar reflections from  chaotic vibrations induced by muscles through recurrence plots and statistical hypothesis testing. We then develop a hierarchical spatial filtering pipeline, at the heart of which is an \textit{entropic ``sieve''}~\cite{chen2007characterization} that eliminates false-positives produced through classical localization algorithms, and identifies regions of high muscle activity.

\sssec{(2) Chaos-Driven Temporal Segmentation: }\textit{Second, during dynamic exercise, limb motion, a dominant source of noise, occurs in the same spatial regions and frequency bands as the muscle vibration itself.} This makes it challenging to segment the signal and identify when specific phases of an exercise take place, in particular the flexing phase, which corresponds to the period of highest muscle activation and force generation. An initial approach is to rely on neural networks for end-to-end segmentation, but this can lead to overfitting as it is challenging to collect a large amount of muscle vibration data from a single person for training. 

Instead, we once again draw from Chaos theory to quantify the vibration signal's \textit{determinism}~\cite{xie2009detection}. To do so, we reconstruct the system's state space using time-delay embeddings, and compute a recurrence matrix that captures the predictability of the signal over time. The determinism of this matrix is high during muscle vibration and drops during periods of postural control, tremor and random motion. We then use this metric to segment the period of the signal contain flexing which is then passed ``hint'' to a downstream neural network, guiding it to focus on relevant regions and learn mappings between the signal and muscle output.


\sssec{(3) Chaos-Informed Muscle State Learning: }\textit{Third, as individuals differ in physiology, variation in muscle fibers, fat thickness, and biomechanical dynamics results in muscle vibrations that can vary drastically across users.} As a result, it can be challenging to create models that can generalize across users. We address this challenge through a dual-model approach with multiple innovations: {\bf (a)} we develop a logistic regression model for fatigue detection that leverages chaos-informed features capturing the nonlinear dynamics of muscle activation; {\bf (b)} we design a neural network for force estimation that incorporates features invariant across individuals, enabling better generalization; {\bf (c)} the neural network models utilize demographic-specific embeddings that encode the user's BMI, age, sex, and other attributes to account for individual differences, and {\bf (d)} we implement a few-shot, human-in-the-loop calibration process that fine-tunes both models to individual users. The calibration process only takes less than a minute in practice (16 seconds in our experiments), and can be integrated into a warm-up routine before exercise, resulting in virtually no additional time cost.\footnote{We note that calibration is standard practice among consumer wearables that perform muscle sensing such as Athos compression apparel~\cite{athos} and Myo armband~\cite{myo}, our approach follows this established industry trend.}

\sssec{Contributions:} Our contributions are as follows:
\squishlist
\item We present the first contactless radar-based system that can estimate muscle output during isotonic and isometric exercise using a chaos theory inspired approach. 
\item In an IRB-approved study with 23 participants, our system outputs percentage of maximum voluntary isometric contraction (MVIC) with RMSE of 5.9\%, which is comparable to a contact-based IMU baseline. It is also able to detect 1-RIR with an AUC of 0.86 and 3-RIR with an AUC of 0.83.
\item We evaluate our system across multiple sessions over a three-week period involving 9 participants and show that our system performance remains largely consistent over time.
\squishend

\section{Primer on muscle physiology}
\label{sec:primer}

\begin{figure*}[ht]
    \centering
    \includegraphics[width=\linewidth]{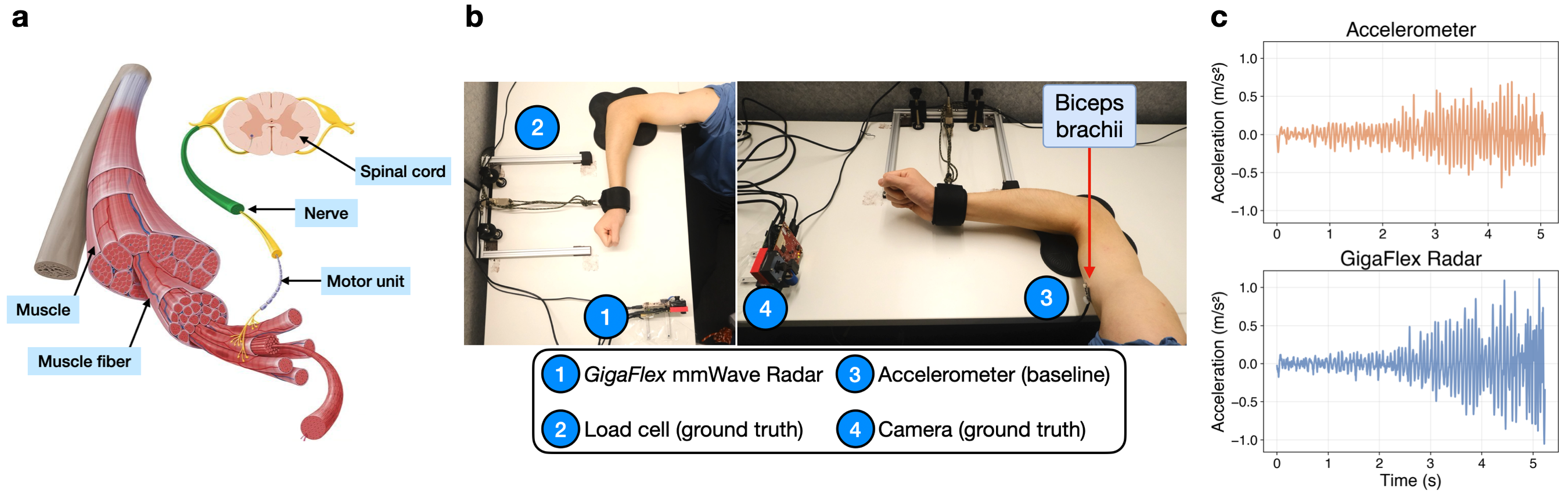}
    \vspace{-2em}
    \caption{{\bf (a)} Neuromuscular origins of surface muscle vibration. {\bf (b)} Experimental setup measuring muscle force output for the isometric (static) exercise of elbow flexion. {\bf (c)} Similarity between muscle vibrations at the bicep measured using a contact-based accelerometer and at a distance of 40~cm using a mmWave radar.}
    \label{fig:anatomy}
    \vspace*{-0.1in}
\end{figure*}

We provide a brief primer on the origin and importance of muscle vibration and fatigue and how muscle output is typically measured.

\subsection{The origins of muscle vibrations}
When a muscle flexes, the brain sends an electrical impulse along the spinal cord to the nerves and fibers. The coordinated firing of thousands of nerves generates the mechanical power necessary for movement and force (Fig.~\ref{fig:anatomy}a). Although, this process appears as a smooth steady motion, such as an arm or leg moving, each muscle fiber produces tiny microscopic twitches that create subtle mechanical displacements at the surface of the skin. 

These firing patterns are asynchronous, occurring at different times and rates as the brain activates different types of fibers to combine fine motor control with rapid power generation. The resulting surface vibrations have distinct characteristics:
\squishenum
\item \textit{Broadband.} They contain low and high frequency components, with slow oscillations from slow-twitch fibers, and faster components from fast-twitch fibers~\cite{islam2013mechanomyogram,matsumoto2025muscle}.
\item \textit{Nonlinear.} The layers of tissue and skin above the muscle distort the vibrations through uneven damping and amplification, resulting in nonlinear transmission of energy to the surface~\cite{orizio1996surface,lubel2023non,beck2005mechanomyographic}.
\item \textit{Pseudo-Stochastic.} As muscle twitches overlap in time, the combined signal appears stochastic and noisy. However, it should be noted that the underlying muscle activities are ultimately \textit{deterministic} because it is governed by the coordinated firing of motor units that follow underlying rule-based dynamics~\cite{orizio1996surface,beck2005mechanomyographic,islam2013mechanomyogram}.
\squishenumend

We show in Fig.~\ref{fig:anatomy}b,c an example recording of the muscle vibrations measured from a contact-based IMU and a radar (setup details in Sec.~\ref{sec:study_design}). Here, we show the IMU acceleration recording from the biceps brachii and the second-derivative of the radar signal to obtain acceleration (details in Sec.~\ref{sec:radar_proc}). The plot shows that the muscle vibrations are distinctly aperiodic in contrast to breathing and heartbeat signals. It also shows that under ideal static conditions the radar signal is similar to that captured from the IMU. We note that the acceleration from the radar signal is slightly higher as it captures reflections over a slightly larger area than the IMU, includes from the body and environment, as well as additional noise introduced during numerical differentiation.

\subsection{The nature of muscle fatigue}

Muscle fatigue is the transient reduction of maximal muscle force output after sustained activation~\cite{enoka_muscle_2008}. In other words, even when the brain sends the same electrical signal, the muscle produces less force due to the weakened efficiency of the muscle fibers. Detecting muscle fatigue is important because once it sets in, individuals often overcompensate with excessive effort, placing strain on the body that can result in injury such as muscle tears and joint instability. It is however challenging to detect fatigue as it can arise from multiple sources of inefficiency from the brain-to-muscle pathway, and lacks a single physiological marker. Furthermore, its expression can vary widely between individuals due to physiological differences.

While scales like the Borg CR-10~\cite{williams2017borg} or rating of perceived exertion (RPE)~\cite{ritchie2012rating} being used to quantify the subjective perception of fatigue, they only correlate moderately to physiological fatigue. Further, these measures are known to vary based on an individual's sense of motivation and pain tolerance~\cite{marcora2009mental,smirmaul2012sense,halperin2020rating}.


\subsection{Measuring muscle force}
Exercises that contract the muscles can broadly be categorized into isometric (static) and isotonic (dynamic) exercises.

\noindent {\bf Isometric (static) exercises.} These involve generating a force without visible movement, such as keeping a dumbbell steady with an outstretched arm~\cite{andersen2010muscle,isometric}. Here, the muscle contracts without changing length to support the weight. Force output is often measured using \textit{maximum voluntary isometric contraction (MVIC)}, the maximum force an individual can exert against a fixed weight. When an individual is not fatigued, their exertion is expressed as a percent of MVIC. When an individual becomes fatigued however, their MVIC decreases~\cite{meldrum2007maximum}. 

MVIC is often measured by having individuals push or pull as hard as they can against a static load cell, to measure a single peak force value~\cite{meldrum2007maximum}. This can be extended into a continuous \textit{force ramp} or ``sweep'', where force increases linearly over time, producing a characteristic curve that  reveals signs of fatigue~\cite{hunter2002task}. Studying muscle output in this context enables more controlled measurements while minimizing confounding effects such as body motion.

\noindent {\bf Isotonic (dynamic) exercises.} These involve rhythmic movement where the muscle alternately shortens and lengthens in a cycle, such as during a bicep curls or squat. Each repetition involves a lifting (concentric) and lowering (eccentric) phase. These exercises are measured using \textit{one-repetition maximum (1RM)} which is the maximum load that can be lifted once for a full cycle~\cite{seo2012reliability,grgic2020test,niewiadomski2008determination}.For example, if a person can complete one full bicep curl with a 5~kg dumbbell and cannot lift a heavier load, that weight represents their maximum load, or 1RM. Training loads are often expressed as a percentage of an individual's 1RM. 
\section{The case for a Chaos-inspired radar}
\label{sec:chaos}

In this section, we aim to show the following:
\squishlist
\item \textit{First}, we illustrate the broadband and pseudo-stochastic nature of muscle vibration signals, and show why conventional time-frequency based filtering methods do not work well. 
\squishend
\squishlist
\item \textit{Second}, we perform statistical analysis demonstrating that the system is nonlinear, which is a key requirement for Chaos theory, and thus motivate why Chaos theory is a good fit for this problem.
\squishend
\squishlist
\item \textit{Finally}, we show how to visualize Chaos in muscle vibrations.
\squishend

\subsection{Limits of conventional filtering}

\begin{figure}[t]
    \centering
    \includegraphics[width=\linewidth]{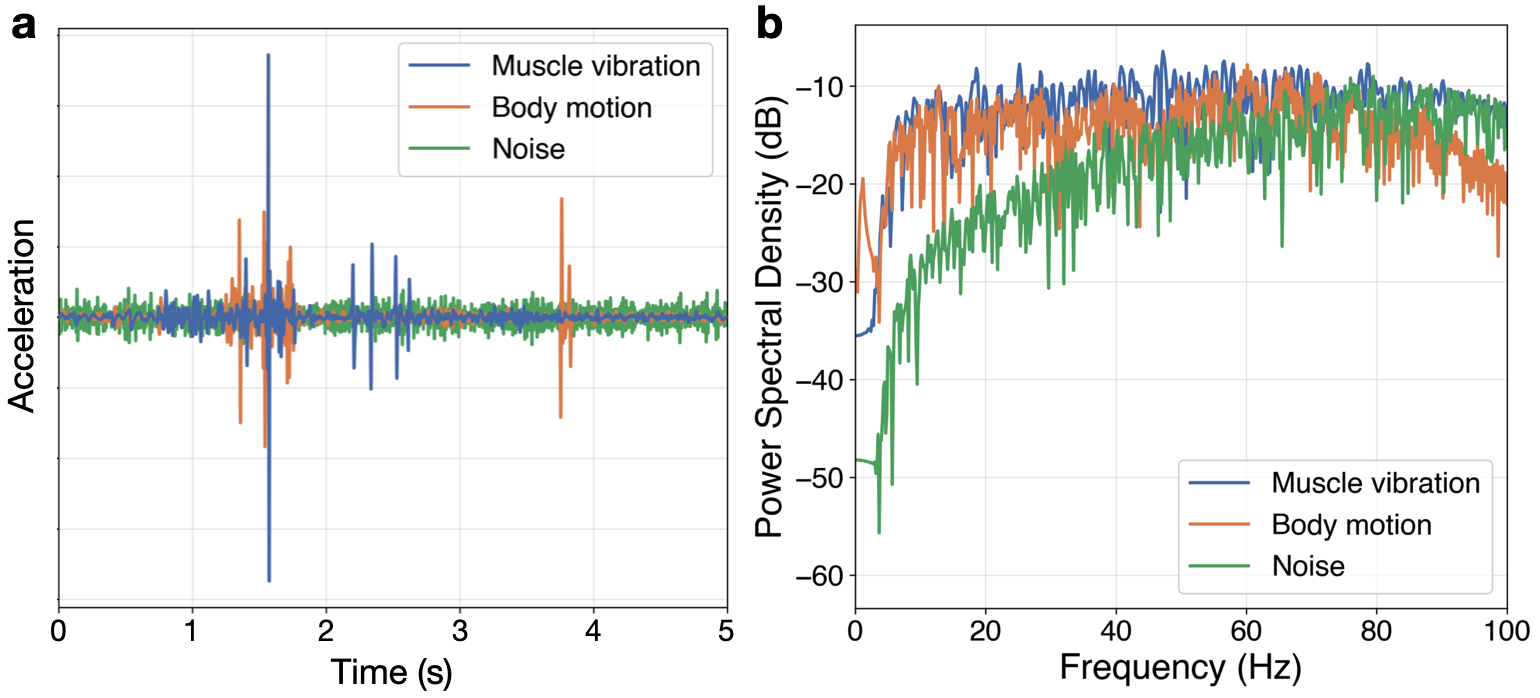}
    \vspace{-1em}
    \caption{Muscle vibrations are broadband and look similar to body motion in the {\bf (a)} time and {\bf (b)} frequency domain.}
    \label{fig:broadband}
\end{figure}

When sensing periodic vital signs in particular breathing and heartbeats, prior knowledge of the signal's characteristic frequency range is typically used to detect its presence and distinguish it from unwanted signals. This is however challenging in the case of muscle vibrations which are broadband (Fig.~\ref{fig:broadband}). As can be seen in the figure the vibrations have a similar profile to body motion, and can be challenging to distinguish. Furthermore, the characteristic frequency range differs from person to person, making it challenging to create a global frequency filter even in perfectly static conditions~\cite{roman2016influence,anders2019inter}.

\subsection{Statistical evidence of Chaos in muscle vibrations}
Given that muscle vibrations are known to be nonlinear while emanating from the deterministic process of motor units firing, we turn to Chaos theory which is designed to model precisely such systems. In fact, prior work~\cite{rodrick2006nonlinear,padmanabhan2004nonlinear,khodadadi2023nonlinear,rampichini2020complexity} has modeled mechanical and electrical muscle vibrations using Chaos theory. By extension, the radar reflections arising from the same underlying muscle vibrations should, in principle, exhibit comparable Chaotic characteristics.

\sssec{Surrogate testing framework.} To determine if this is indeed the case, we leverage a surrogate data testing framework~\cite{schreiber1996improved} which is a statistical testing framework that distinguishes nonlinear Chaotic systems to linear ones. The high level idea is to generate surrogate signals of the muscle vibrations that retain its linear properties. In other words, it maintains the same frequency spectrum and amplitude distribution, but does not have any of its nonlinear properties, should there be any that exist in the signal.  If measures of nonlinearity differ significantly between the signal under test and these surrogates, then the original signal is deemed to contain genuine chaotic dynamics.

\begin{table}[htbp]
\centering
\small
\caption{Statistical testing of nonlinearity in signals. $p$-values below 0.05 indicate rejection of the null hypothesis that the signal is linear, and supports the hypothesis of deterministic nonlinear dynamics. $^{**}$ means $p < 0.01$.}
\label{tab:surrogate_results}
\vspace{-1em}
\begin{tabular}{lcc}
\toprule
\textbf{Signal Type} & \textbf{Sample Entropy} & \textbf{Correlation Sum} \\
& \textbf{($p$-value)} & \textbf{($p$-value)} \\
\midrule
\multicolumn{3}{l}{\textit{Simulation: Linear Stochastic}} \\
\quad White Noise & 0.25 & 0.27 \\
\quad Autoregression (AR) & 0.25 & 0.41 \\
\midrule
\multicolumn{3}{l}{\textit{Simulation: Known Chaos}} \\
\quad Lorenz Attractor~\cite{tucker1999lorenz} & $^{**}$ & $^{**}$ \\
\midrule
\multicolumn{3}{l}{\textit{Real-world: Physiological}} \\
\quad Muscle Vibration (60\% MVIC) & $^{**}$ & $^{**}$ \\
\quad Body Motion & $^{**}$ & $^{**}$ \\
\midrule
\multicolumn{3}{l}{\textit{Real-world: Environmental}} \\
\quad Noise & 0.08 & 0.17 \\
\bottomrule
\end{tabular}
\end{table}


\begin{figure*}[h]
\centering
\includegraphics[width=\linewidth]{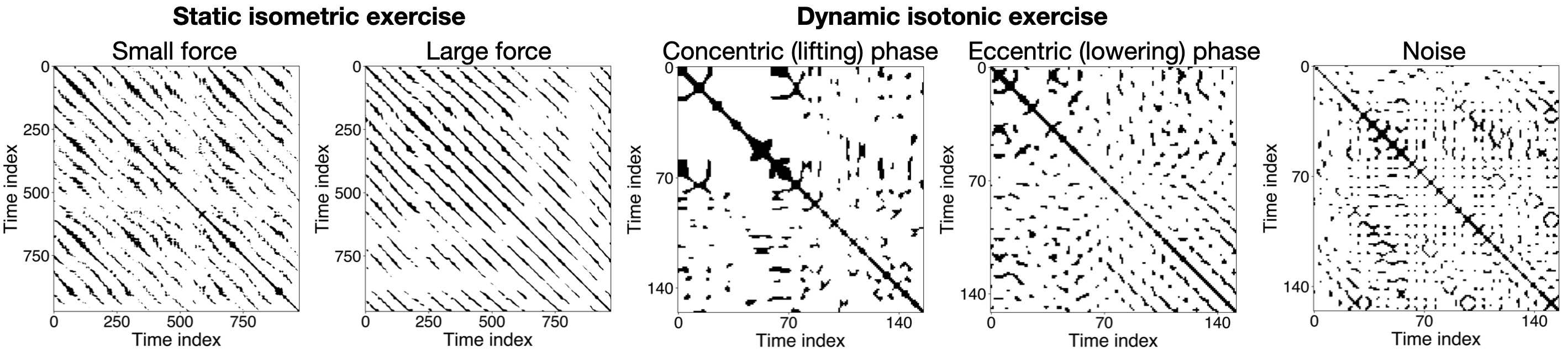}
\vspace{-2em}
\caption{Recurrence plots for different scenarios: static isometric exercise, dynamic isotonic exercise, and noise. These plots show how similar a muscle signal is to itself over time. Dark cells indicate when there is a repeated pattern.}
\label{fig:rp_examples}
\end{figure*}

\noindent \sssec{Testing realworld and simulation signals.} 
We ran our surrogate data testing framework on both simulated and real-world signals using 100 surrogate signals using the Iterative Amplitude Adjusted Fourier Transform method~\cite{schreiber1996improved}, and quantify the level of chaos using \textit{sample entropy} which characterizes the irregularity of the signal~\cite{richman2004sample}, and \textit{correlation sum} which characterizes the density of the signal's recurrence matrix (Sec.~\ref{sec:viz})~\cite{grassberger1983measuring}. We summarize our results in Table~\ref{tab:surrogate_results} below:

\squishlist
\item Simulations of linear signals (white noise and AR(2)) showed that their behavior was not significantly different from the linear surrogates, suggesting no strong evidence of chaos.
\item A simulation of the chaotic Lorenz attractor~\cite{tucker1999lorenz} had $p$-value < 0.01, rejecting the null hypothesis of linearity, which is expected.
\item Real-world signals of muscle vibrations and body motion had $p$-value < 0.01. We reject the null hypothesis and show they exhibit deterministic nonlinear dynamics consistent with chaotic behavior. \textit{Using measures of Chaos we can heuristically distinguish between these two types of signals for localization and segmentation.}
\item Environmental noise had $p$-value 0.08, which does not meet the significance threshold. We fail to reject the null hypothesis and show that the signal is consistent with a linear stochastic process.
\squishend

In summary, this statistical analysis shows that the muscle vibrations as measured by the radar do exhibit nonlinear characteristics and motivate the need to use Chaos theory as opposed to conventional linear models which would not be able to capture the nonlinear dynamics of the signal.

\subsection{Visualizing Chaos}
\label{sec:viz}

While statistical tests are useful to provide quantitative assurance the muscle vibrations are indeed Chaotic, it would be useful to have a visual method to illustrate these dynamics.

An established technique of visualizing the Chaos within signals is to plot its \textit{recurrence} which captures the deterministic nonlinear dynamics of chaotic systems~\cite{marwan2007recurrence}. Intuitively, recurrence captures when a system revisits a similar \textit{state}, and a recurrence plot is a symmetrical $N \times N$ time-time mapping of those revisits, where each point in the array encodes whether two states are close in the state space. These recurrence plots not only visualize Chaos but are also a key primitive that underlies the techniques used as part of our Chaos-inspired radar architecture.

\noindent \subsubsection{Understanding recurrence plots.} Fig.~\ref{fig:rp_examples} shows the recurrence plots of the muscle vibrations measured by the radar for static isometric and dynamic isotonoic exercises, as well as noise. The figure show the following:
 \squishlist
\item Black regions in the plot are indicative of similar vibration patterns, while white regions indicate different or no regularity.
\item \textit{Static isometric exercises.} The clear diagonal lines represent regular, repeating vibration patterns, which is expected since the muscle length remains constant and the contraction is steady under controlled conditions.
\item \textit{Larger isometric forces.} Results in denser and more continuous diagonal stripes indicating vibration patterns that repeat more.
\item \textit{Dynamic isotonic exercises.} The muscle shortens and lengthens with each cycle, so in contrast, the patterns are more broken and are less uniform.
\item \textit{Concentric (lifting) phase} shows stronger and denser  patterns vs. \textit{eccentric (lowering) phase} which has more scattered patterns.
\item \textit{Noise.} Random scattered patterns with less geometric structure.
 \squishend

\section{{\sysname} architecture}

\subsection{Bringing Chaos theory to radar sensing}
Now that we have tested for chaos in muscle vibrations and visualized the chaos, the next question is \textit{how} techniques from Chaos theory can be leveraged to sense these vibration signals effectively. 



In this section we describe our approach to weaving different techniques from Chaos theory into the classical radar sensing and learning architecture to address the core technical challenges outlined in the Introduction to obtain the muscle vibration signals. Specifically, we describe:

\squishenum
\item A hierarchical two-stage spatial filtering pipeline using an \textit{entropic ``sieve''} to guide the localization of the muscle vibration 
\item A temporal segmentation algorithm to extract the period of maximal muscle activation using \textit{determinism}.
\item A learning pipeline to adapt to differences across individual physiology using \textit{chaos features and demographic embeddings}.
\squishenumend


\subsection{Basics of radar-based muscle vibration sensing}
\label{sec:radar_proc}

\begin{figure*}[h]
    \centering
    \includegraphics[width=\linewidth]{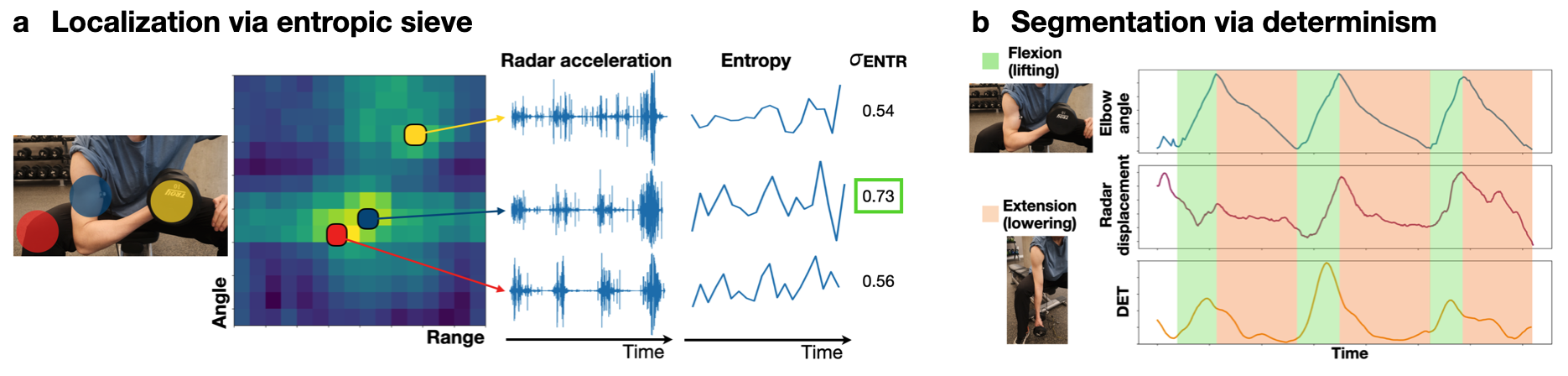}
    \vspace{-2em}
    \caption{{\bf (a) Muscle localization via entropy.} Classical radar localization using CFAR produces false positives from body movement (\protect\tikz[baseline=-0.5ex]\protect\node[circle,draw=black,fill=myred,inner sep=2pt, line width=.5pt] {}; knee movement) as well as motion in the environment (\protect\tikz[baseline=-0.5ex]\protect\node[circle,draw=black,fill=myyellow,inner sep=2pt, line width=.5pt] {}; dumbbell motion), and can only be used as a coarse-grained filter. The Chaos dynamics of muscle vibrations can be characterized using an entropic ``sieve'', $\sigma_{ENTR}$, where values below a certain threshold can be discarded, and the maximum value can be regarded as the muscle vibration signal. {\bf (b) Segmentation of lifting phase via determinism.} Relying on radar displacement to determine the start and end of a lift is unreliable because the signal is often contaminated by small tremors and subtle body motions. In contrast, determinism (DET) metric captures the temporal regularity of muscle vibrations, enabling more precise segmentation of the lifting phase.}
    \label{fig:cfar}
\end{figure*}

We first begin with a brief overview of how radar sensing works to measure the muscle vibration under idealized conditions (Sec.~\ref{sec:radar_proc}).



\subsubsection{Measuring skin displacement}
Using a mmWave radar, our system transmits frequency-modulated continuous wave (FMCW) signals to capture the subtle skin surface vibrations induced by muscle contractions. It transmits at 77~GHz with a frequency slope of 63.343~MHz/$\mu$s.  As minute displacement variations at the skin surface change the round-trip propagation distance $d(t)$ between the radar and the skin~\cite{alizadeh_mmWaveBodyPenetration_2019}, we analyze variations in the received signal's phase $\phi(t)$ to extract the muscle vibration signal, given by:
\begin{equation}
\phi(t) = \frac{4\pi d(t)}{\lambda} + \phi_0
\end{equation}

\noindent where $\lambda$ represents the wavelength of the transmitted signal and $\phi_0$ denotes the initial phase offset. 




We look at the acceleration of the signal instead of the displacement to emphasize the signal changes associated with the muscle vibrations. The radar phase signals first undergo bandpass filtering (5--100~Hz, 5th-order Butterworth) to isolate the muscle vibration band while removing DC drift and high-frequency noise. We then obtain the acceleration using a numerical finite difference method for the second derivative~\cite{zhao_emotion_2016,diff}:
{\footnotesize
\begin{equation}
d''(t_0) = 
\frac{
4d(t_0)
+ [d(t_{1}) + d(t_{-1})]
- 2[d(t_{2}) + d(t_{-2})]
- [d(t_{3}) + d(t_{-3})]
}{
16\,(\Delta t)^2
}
\end{equation}
}
\noindent where $\Delta t$ represents the sampling interval, and 
$d(t_{i})$ denotes the displacement $i$ samples away. This preprocessing pipeline enhances the underlying system dynamics for chaos analysis by removing noise components that can obscure the deterministic muscle vibration patterns. Since chaotic dynamics are sensitive to noise, filtering the signal to retain only the physiologically relevant frequency band strengthens the recurrence structures and improves the reliability of nonlinear features extracted from the muscle vibration system~\cite{wendi2018extended}.




\subsubsection{Isolating the muscle vibration using spatial filtering} On our radar, we use 2 transmitter and 4 receive antennas, forming a virtual array of 8 antennas that enables beamforming towards the region of muscle vibration. To isolate the signal, we employ range-angle processing by applying a two-dimensional Fast Fourier Transform (2D-FFT) on the raw radar data. First, a range FFT $\mathcal{F}$ is applied along the fast-time dimension to distinguish reflections from different distances. We obtain the frequency components:
\begin{equation}
X_n = \mathcal{F}[x[t]]
\end{equation}
where $x[t]$ represents the discrete ADC samples and $n$ is the range bin index from 0 to $N$. Next, an FFT is applied across the antennas to separate reflections along the angular dimension. Zero-padding is applied to enhance angular resolution:
\begin{equation}
X_{n,\theta}=\mathcal{F}[X_n[r]]
\end{equation}
where $\theta$ represents the azimuth angle bin index and $r$ is the receiving antenna index. This process spatially decomposes the radar's point of view, allowing the signal to be measured at distinct range and angle bins. However, selecting the range of bins that correspond to the target muscle region is challenging given the broadband nature of the signal, and the presence of body motion during dynamic exercise. To design an appropriate bin selection strategy, we characterize the muscle vibration signal in this idealized setting using Chaos theory as described in the next subsection.


\subsection{Localization with an entropic sieve}
\label{sec:loc}

In this section, our goal is to find the spatial region with the muscle vibration. To do this, we present a hierarchical two-stage pipeline to perform spatial filtering and intelligently select the region containing the muscle signal. The first stage uses, Constant False Alarm Rate (CFAR), a classical radar localization algorithm as a coarse-grained filter. We show why this technique alone is not sufficient at localizing the muscle signal. The second stage uses a Chaos-based metric of signal regularity known as entropy (ENTR). The metric's sensitivity to signal regularity allows it to distinguish muscle activation by quantifying the change in variance caused by varying contraction lengths and torque directions.

\noindent {\bf Stage 1. Constant False Alarm Rate (CFAR)} is a widely used algorithm~\cite{richards2014radar} for automatic spatial selection of targets using radar in cluttered environments. The idea is to maintain a constant probability of false alarms despite varying noise. It does this by dynamically estimating noise power in a window around each range-angle bin, and only selects that bin as containing the muscle signal if it exceeds an adaptive threshold. Specifically, a cell under test (CUT) with power $P_{\text{CUT}}$ is declared as a detection if:
\begin{equation}
P_{\text{CUT}} > \alpha \, P_n
\end{equation}

\noindent where $Pn = \frac{1}{N_T} \sum_{i=1}^{N_T} P_i$ is the average noise power estimated from $N_T$ training cells surrounding the CUT (excluding $N_G$ guard cells), and the threshold scaling factor $\alpha$ is determined from the desired false alarm rate $P_{\text{fa}}$ as:
\begin{equation}
\alpha = N_T \left( P_{\text{fa}}^{-1/N_T} - 1 \right)
\end{equation}

Here, we apply a two-dimensional Cell Averaging Constant False Alarm Rate (2D CA-CFAR) detector with training cells $(Tr,Ta)=(3,1)$, guard cells $(Gr,Ga)=(1,1)$, and probability of false alarm $P_{\text{fa}}=0.08$. We show in Fig.~\ref{fig:cfar}a that the algorithm highlights regions of higher reflections. While this includes the region of the muscle, it also contains reflections from different regions of the body such as the knee and the environment. This is because CFAR relies only on the power of the signal reflection but does not leverage other features of the signal to distinguish the muscle from other objects.

The key challenge, however, lies in the trade-off between sensitivity and specificity: increasing the threshold for signal detection to suppress false positives also causes the subtle muscle vibration to fall below the detection limit. This trade-off motivates the need for a second stage that moves beyond energy detection and instead exploits the Chaos characteristics of the signal for detection.

\noindent {\bf Stage 2. Entropic Sieve.} Next, we use entropy (ENTR)~\cite{marwan2007recurrence} to characterize the time-series signal in each range-angle bin. Specifically, entropy is used to characterize if the signal in a spatial region exhibits a relatively stationary and complex signal versus static reflectors from other parts of the body or dynamic motion from the environment that do not, and is defined as:
\begin{equation}
\text{ENTR} = -\sum_{l=l_{min}}^{N} p(l) \ln p(l)
\label{eq:entropy}
\end{equation}
where $p(l)$ denotes the normalized probability distribution of diagonal line lengths $l$ in the recurrence plot, and $l_{\min}$ is the minimum diagonal length considered. 

In our design, we look at non-overlapping 1-second windows over a 10-second period and look at the standard deviation of the 10 entropy values, $\sigma_{\text{ENTR}}$, for each range-angle bin. The bin within the top 5 $\sigma_{\text{ENTR}}$ that has the highest signal strength is then selected as the muscle region of interest. 

We note that in the context of the static isometric exercise of pulling on the load cell, the muscle region of interest does not change. While in the dynamic isotonic exercises evaluated in our paper, the muscle region of interest, specifically the bicep area, remains relatively still, while the forearm moves dynamically. In our paper, we focus on analyzing the muscle region of the static forearm. During our experiments, we only need to perform a one-time calibration per set of bicep curls to localize the muscle region. 

\subsection{Determinism-driven signal segmentation}
\label{sec:seg}



Now that we have identified the spatial region containing the muscle region, our goal is to segment the time period when the muscle is of highest activation. In the case of static isometric exercises, this is simply when the user starts tensing their muscles. In the case of rhythmic dynamic isotonic exercises with lifting (concentric) and lowering (eccentric) cycles, and sometimes holding in between, our goal is to isolate the concentric phase. The concentric phase contains increasing amounts of vibration as it produces more force, while during the eccentric stage the muscle generates more vibration impulses as it passively releases the weight, controlling the downward speed. During the holding stage, there is random unwanted tremor as the arm tries to balance the weight.

We illustrate why this is challenging in Fig.~\ref{fig:cfar}b, where the user performs a full bicep curl cycle with the a ground truth measure of the elbow angle as measured using the camera. The elbow angle is reconstructed using the ViT-Pose reconstruction library~\cite{xu2022vitpose} and angle computed. When analyzing the radar reflection we note that the peaks of the signal do not line up with the ground truth angle. This is because the signal is contaminated with minor body motion such as low-amplitude tremors and shaking of the bicep, although it the body remains relatively still from a macro perspective. 

To address this, we again turn to Chaos theory which has been used to analyze other time-series biosignals including electrical electrocardiogram (ECG) and optical photoplethysmogram (PPG) signals for recurrency~\cite{naschitz2004patterns,marwan2002recurrence}. We use the \textit{determinism (DET)}~\cite{webber1994dynamical} metric which quantifies the amount of temporal regularity in the system and can be used to distinguish between different events in a signal over time and is defined as:
\begin{equation}
\text{DET} = \frac{\sum_{l=l_{min}}^{N} l \cdot P(l)}{\sum_{i,j} \mathbf{R}_{i,j}}
\label{eq:det}
\end{equation}
where $R_{i,j}$ is the binary recurrence matrix, $P(l)$ is the count of diagonal lines of length $l$, and $l_{min}$ is the minimum line length included.We apply this to the selected range-angle bin, and compute DET using a window size of 1-second, with a moving window with overlap of 0.1~s. We show in Fig.~\ref{fig:cfar}b that the DET metric follows the same trend of the elbow angle through the concentric phase. 


To precisely segment the concentric phase from the continuous muscle activity signal, we first apply a \texttt{savgol\_filter} (window length 0.15\,s, polynomial order 3) to smooth the signal. We then use peak detection on the DET metric to identify potential repetition boundaries, followed by tangent-based foot-point estimation to determine the exact starting point of each concentric phase. Specifically, we fit a linear regression model to a 10-sample window on the ascending slope of each peak and extrapolate it to its intersection with the local minimum value, marking the precise onset of the concentric phase. This method provides robust segmentation, as the tangent-based approach captures the overall trend of muscle activation rather than being sensitive to local fluctuations.



\subsection{Chaos-informed learning of muscle state}
\label{sec:learn}

\begin{figure}[t]
    \centering
    \includegraphics[width=0.4\textwidth]{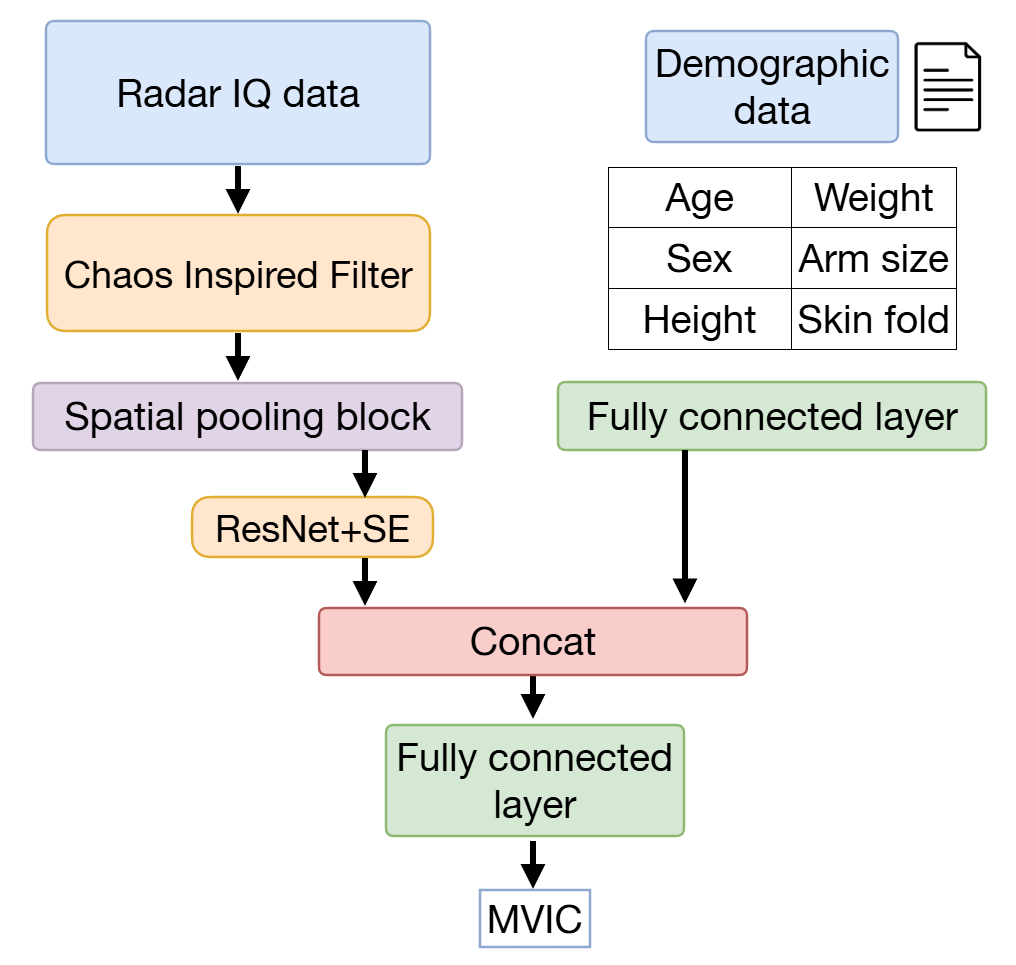}
    \vspace{-1em}
    \caption{{\bf Muscle force estimation network architecture. }}
    \label{fig:ml}
\end{figure}

Now that we have located and segmented the signal to isolate muscle response patterns, our goal is to estimate MVIC levels to accurately quantify the user's exerted effort and detect muscle fatigue.

\subsubsection{Limits of conventional force estimation methods.} 
\begin{figure}[t]
    \centering
    \includegraphics[width=0.4\textwidth]{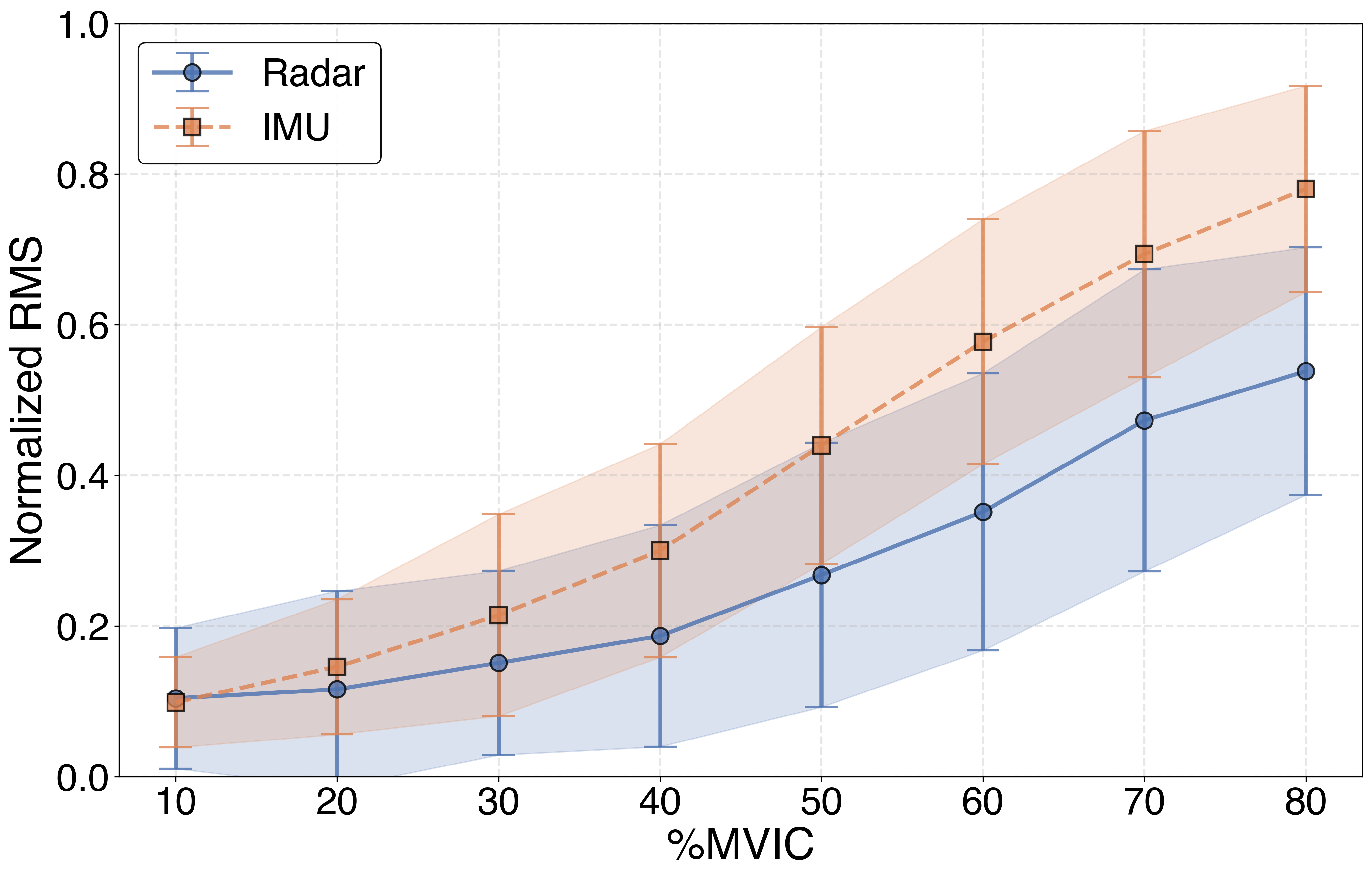}
    \vspace{-1em}
    \caption{{\bf Conventional amplitude based force estimation methods exhibits substantial overlapping error bars across force levels for both radar and IMU, motivating the need for a learning-based approach.}}
    \label{fig:rms_error}
\end{figure}

Conventional methods for muscle force estimation, rely on a signal root mean square (RMS) amplitude value as a coarse-grained measure of force. Fig.~\ref{fig:rms_error} shows that RMS increases as a function of MVIC\% for the radar and IMU data collected across all subjects in our study from MVIC levels of 10 to 80\% in increments of 10\%. However, the error bars between different MVIC levels overlap. We quantify this by computing the overlap percentage between error bar ranges (mean ± standard deviation) for two adjacent MVIC levels with ranges $[\mu_1 - \sigma_1, \mu_1 + \sigma_1]$ and $[\mu_2 - \sigma_2, \mu_2 + \sigma_2]$. The overlap percentage is calculated as: 
\begin{equation} \footnotesize
Overlap(\%) = \frac{max(0, min(\mu_1 + \sigma_1, \mu_2 + \sigma_2) - max(\mu_1 - \sigma_1, \mu_2 - \sigma_2))}{max(\mu_1 + \sigma_1, \mu_2 + \sigma_2) - min(\mu_1 - \sigma_1, \mu_2 - \sigma_2)} \times 100
\end{equation}

Radar RMS measurements showed an average overlap of 67\%, while IMU RMS measurements exhibited 48.1\% average overlap indicating substantial overlap. This analysis shows simple statistical features cannot reliably discriminate between different force levels, motivating the need for a learning-based approach.

\subsubsection{Learning-based force estimation.} Below we describe key features of our learning-based approach.


\noindent \textbf{Spatial aggregation of muscle vibrations.} As muscle vibrations ripple across multiple spatial bins we use the 9 adjacent bins around the bin selected by our Chaos-inspired localization algorithm as input to our model. The challenge however is that neighboring bins also introduce additional noise. To suppress this noise, we apply a spatial pooling block~\cite{zhu2025measuring} that reduces the dimensionality from 9 bins to 3 bins.

\noindent \textbf{Modeling fast and slow twitch vibrations.} Inspired by U-Net's effectiveness at time-series feature extraction, we adapt its encoder architecture for our muscle vibration sensing system. As muscles have slow and fast-twitch fibers that give rise to vibrations of different amplitudes, we design our architecture to have a hierarchy of convolutional branches that capture fine-grained rapid muscle fiber vibration and coarse-grained force patterns using different kernel sizes. To reduce the effect of noise, we leverage channel-wise attention using Squeeze-and-Excitation (SE) blocks~\cite{hu2018squeeze} which are trained to focus on muscle activation patterns.


\noindent \textbf{Raw I/Q signals.} Phase signals suffer from instability near zero-crossing with environmental noise, causing the second derivative of radar phase to exhibit clutters that obscure underlying muscle vibration patterns. Due to this, we use raw I/Q channels as input features rather than the derived phase signals. 

\subsubsection{Addressing cross-subject variance.} Individual differences in muscle composition, mass, and biomechanics make it challenging for models to generalize to unseen individuals. To address this challenge, we leverage three different strategies:

\noindent {\bf Few-shot calibration protocol.} For each new subject, we record their maximum voluntary isometric contraction (MVIC), and 2 seconds of isometric contractions at force levels ranging from 10--80\% MVIC in 10\% increments. This brief 16-second calibration can easily be incorporated into a normal warm-up routine. The resulting dataset is then used to fine-tune the pre-trained model for that individual.

\noindent {\bf Signal normalization.} Using the calibration data, we compute subject-specific mean and standard deviation values to perform $z$-score normalization on subsequent measurements. This normalization removes baseline amplitude offsets between subjects.

\noindent {\bf Demographic embeddings.} Finally,  we incorporate subject-specific demographics (age, sex, height, weight, BMI, skin fold thickness, gym-goer status) as an embedding into the model. This is important given that some attributes such as body-mass index (BMI) can play a significant role in the shape of the muscle outputs.

\subsubsection{Regression formulation.} 
Although our protocol instructs subjects to maintain a constant force at MVIC levels at multiples of 10\%, in reality it is challenging to maintain a perfectly consistent muscle force output. Given this reality, we formulate the problem of muscle output prediction to be a continuous regression problem, where we use the mean normalized load-cell force over each 0.4~s radar window as ground truth.


\subsubsection{Fatigue Detection with Chaos Features}
For static, isometric contractions, fatigue is considered to have occurred when MVIC drops by 30\%~\cite{Walton2002}. For dynamic, isotonic contractions, we quantify fatigue using the Repetitions in Reserve (RIR) metric~\cite{Refalo2023}, where 1-RIR corresponds to near maximal exertion and imminent muscle failure, while 3-RIR corresponds to high intensity effort and approaching muscle failure. We note that because it takes 24--48 hours of recovery from fatigue, participants are only able to be tested once for fatigue in a particular session.

In our system, we leverage Chaos features extracted from the recurrence plot including entropy ($ENTR$), vertical line entropy ($ENTR_v$), and DFA scaling exponents, which are known to be invariant to fluctuations in vibration amplitude and small changes in orientation. These features are then input into a logistic regression which we find is sufficient at producing performance results comparable to an IMU baseline.

\subsubsection {Training methodology.} For the force estimation model, we employ a two-stage pipeline. The model is pre-trained across multiple subjects using the AdamW optimizer (learning rate $5 \times 10^{-4}$, weight decay 0.03, Cosine Annealing scheduler) for 100 epochs on 0.4-second segments (400 samples at 1~kHz). This window size was selected as muscle vibrations do not fall below 5 Hz (0.2-second segments)~\cite{orizio1996surface}.



\section{Study design}
\label{sec:study_design}

\begin{figure*}[t]
    \centering
    \includegraphics[width=\textwidth]{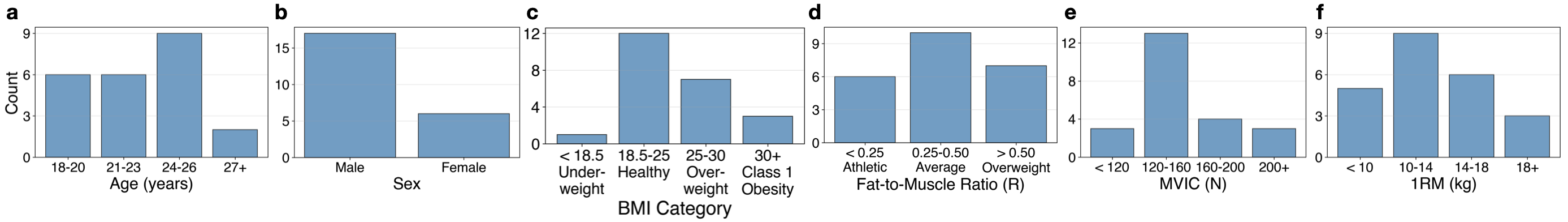}
    \vspace{-2.5em}
    \caption{Demographic summary of participants in the human subjects study.}
    \label{fig:demos}
\end{figure*}

\subsection{Hardware setup}


Our radar-based system uses a Texas Instruments AWR1843BOOST millimeter-wave radar~\cite{ti_radar} configured at 512 samples per chirp at a 1000~Hz frame rate. The radar is positioned 40~cm from the biceps (Fig.~\ref{fig:anatomy}b). Our setup includes the following sensors:
\squishlist
\item {\bf Ground truth reference of muscle output.} A load cell provides a measure of muscle output during the isometric exercise of static elbow flexion. The load cell uses a 100~kg S-type tension sensor interfaced with an ADS1256 24-bit ADC~\cite{adc} connected to an Arduino R4 Minima~\cite{arduino_r4}, that samples at 1000~Hz. A rope connects the load cell to a wrist band, enabling controlled force application during isometric exercises (Fig.~\ref{fig:anatomy}b).

\item {\bf Baseline mechanomyography measurements.} The IMUs (LSM6DSOX~\cite{imu}) are positioned at the short head of the biceps brachii using medical tape. The IMUs are connected to an Arduino DUE~\cite{due}, that samples at 1000~Hz. \textit{The placement of the IMU do not interfere with the radar measurements. This is because the muscle vibrations propagate across the entire muscle area, and the radar uses reflections across multiple regions for processing.}

\item {\bf Ground truth reference of muscle location.} This is captured using a camera (Arducam OV9782~\cite{arducam}) with 100-degree field-of-view lens. The camera is co-located with the radar and assesses the performance of our radar-based muscle localization algorithm.

\squishend

\subsection{Participant demographics}

We conducted a human subjects study at our university campus that was approved by our Institutional Review Board (Protocol \#STUDY2025\_00000178). For all participants, written informed consent was obtained. We recruited 23 participants (Fig.~\ref{fig:demos}) with a mean age of $23~\pm~3$ years, the mean height was $175~\pm~8~\mathrm{cm}$, and weight was $78~\pm~15~\mathrm{kg}$. From these measurements, we also computed body mass index (BMI)~\cite{bmi}, which was on average $25~\pm~4~\mathrm{kg/m^2}$. Using standard anthropometric equations~\cite{frisancho1990anthropometric,heymsfield1982anthropometric}, we computed fat-to-muscle ratio which was on average $0.4 \pm 0.2$. 



\subsection{Experimental protocol}

All participants rested for at least three days prior to their first session to ensure full muscular recovery. Subjects completed a standardized warm-up and familiarization session with the apparatus.

\noindent The experimental protocol is divided into three phases:
\squishenum

\item {\bf Isometric (Static) Force Estimation.} This was evaluated through elbow flexion with a load cell. We first conduct a baseline measurement of MVIC (Sec.~\ref{sec:primer}) by having participants pull at the load cell as hard as they can. Next, we measure a {continuous force profile}~\cite{orizio_surface_2003} by having them linearly increase their muscle exertion from 0 to 80\% over 6.75~s using the visual feedback of load cell output as a guide. Finally, we record the muscle vibrations when the participant pulls and holds at the load cell at discretized MVIC levels from 10--80\% in increments of 10\%, at randomized levels.

\item {\bf Isotonic (Dynamic) Force Estimation.} This was evaluated through concentric bicep curls. We find the participant's 1RM by providing them with progressively heavier dumbbells and instructing them to perform bicep curls at a controlled tempo consistent with prior  protocols~\cite{androulakis_korakakis_optimizing_2024}. Following this, participants performed three repetitions of bicep curls at four load levels corresponding to 10, 30, 50, and 70\% of their 1RM, with 2-minute rest intervals between load conditions.




\item {\bf Detecting Muscle Fatigue.} Using established protocols from prior biomechanical literature~\cite{enoka_muscle_2008, allen_skeletal_2008}, we induced fatigue by having participants perform concentric bicep curls at an intensity of 40\% 1RM at a frequency of 6~s per repetition until the they could no longer complete a full repetition. An MVIC strength measurement and continuous force profile is taken before and after this procedure as a measure of pre- and post-fatigue muscle output levels.

\squishenumend

To assess system performance over time, 9 participants repeated the protocol over three weeks, with one-week intervals between sessions.


\section{Evaluation}

We evaluate our system's performance at force estimation for isometric (static) and isotonic (dynamic) exercises, as well as fatigue detection, finally we perform benchmark experiments evaluating different aspects of the system pipeline, and across different experimental scenarios.

\begin{figure*}[t]
    \centering
    \includegraphics[width=\textwidth]{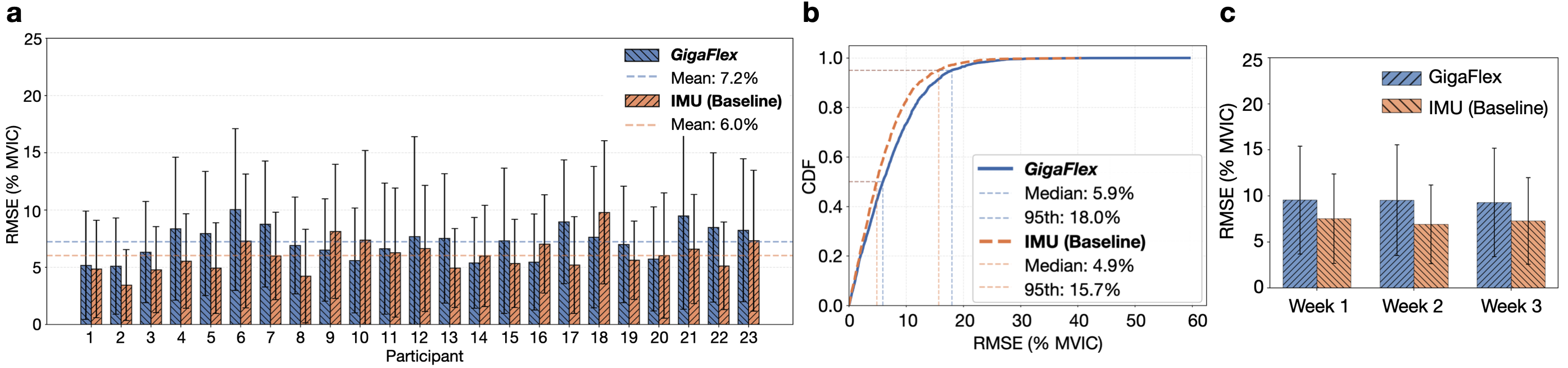}
    \vspace{-2em}
    \caption{{\bf Isometric (static) force evaluation results.} Single-session MVIC errors comparing our contactless system with a contact-based IMU baseline shown for {\bf (a)} individual participants {\bf (b)} CDF across participants in study. {\bf (c)} Cross-session MVIC errors over a three week period demonstrating similar performance over time.}
    \label{fig:isometric}
\end{figure*}

\subsection{Isometric (static) force estimation}
\label{sec:eval_force}

We evaluate isometric force estimation accuracy using data collected during elbow flexion exercises, with load cell measurements serving as ground truth labels. We use Leave-One-Subject-Out cross-validation with fine-tuning. We train on all but one held-out subject, fine-tune on a 2-second MVIC calibration segment per force level for 20 epochs, and evaluate using 2-second windows. This is repeated for all subjects. This methodology is also used to compute baseline results from the IMU data. We use the z-axis of the IMU because it is along the axis of vibration. 


\subsubsection{Single-Session}

Fig.~\ref{fig:isometric}a presents MVIC estimation root-mean-square error (RMSE) across all 23 participants, comparing our radar-based system against an IMU-based baseline. We note that the use of IMUs to measure muscle vibrations via mechanomyography (MMG) is a widely used approach in the space of muscle force output sensing. Our contactless radar approach achieves comparable performance to contact-based IMU sensors, with RMSE of 7.2\% and 6.0\% MVIC, respectively. 


\noindent {\bf Subgroup analysis.} We perform a subgroup analysis across BMI and find that individuals who are in the BMI categories of underweight/normal had an MVIC error of 9.4\%, while for those in the overweight/obese category, their MVIC error was 9.5\%, suggesting that our system works comparably across individuals of different compositions. 


\noindent\textbf{Error Distribution Analysis.} Fig.~\ref{fig:isometric}b shows the cumulative distribution function (CDF) of force estimation errors. Our system achieves a median error of 5.9\% MVIC, with the 95\% percentile of 18.0\% MVIC. This is similar to the IMU baseline with median error of 4.9\% and 95th percentile error of 15.7\%.



\subsubsection{Performance across time}


To assess performance over time, nine participants completed three sessions one week apart. For each leave-one-subject-out run, we trained on the remaining participants and calibrated on that subject’s data for each day. Fig.~\ref{fig:isometric}c shows performance over time with the MVIC error at 9.5\% in Week 1, 9.5\% in Week 2, and 9.3\% in Week 3. The modest variation of 0.2\% over three weeks suggests that our system's algorithms are able to adapt to changes across different session in particular small changes in sensor positioning, as well as physiological changes such as hydration status and baseline muscle state at the start of a session. 





\subsection{Isotonic (dynamic) force estimation}

\begin{figure}[t]
    \centering
    \includegraphics[width=.3\textwidth]{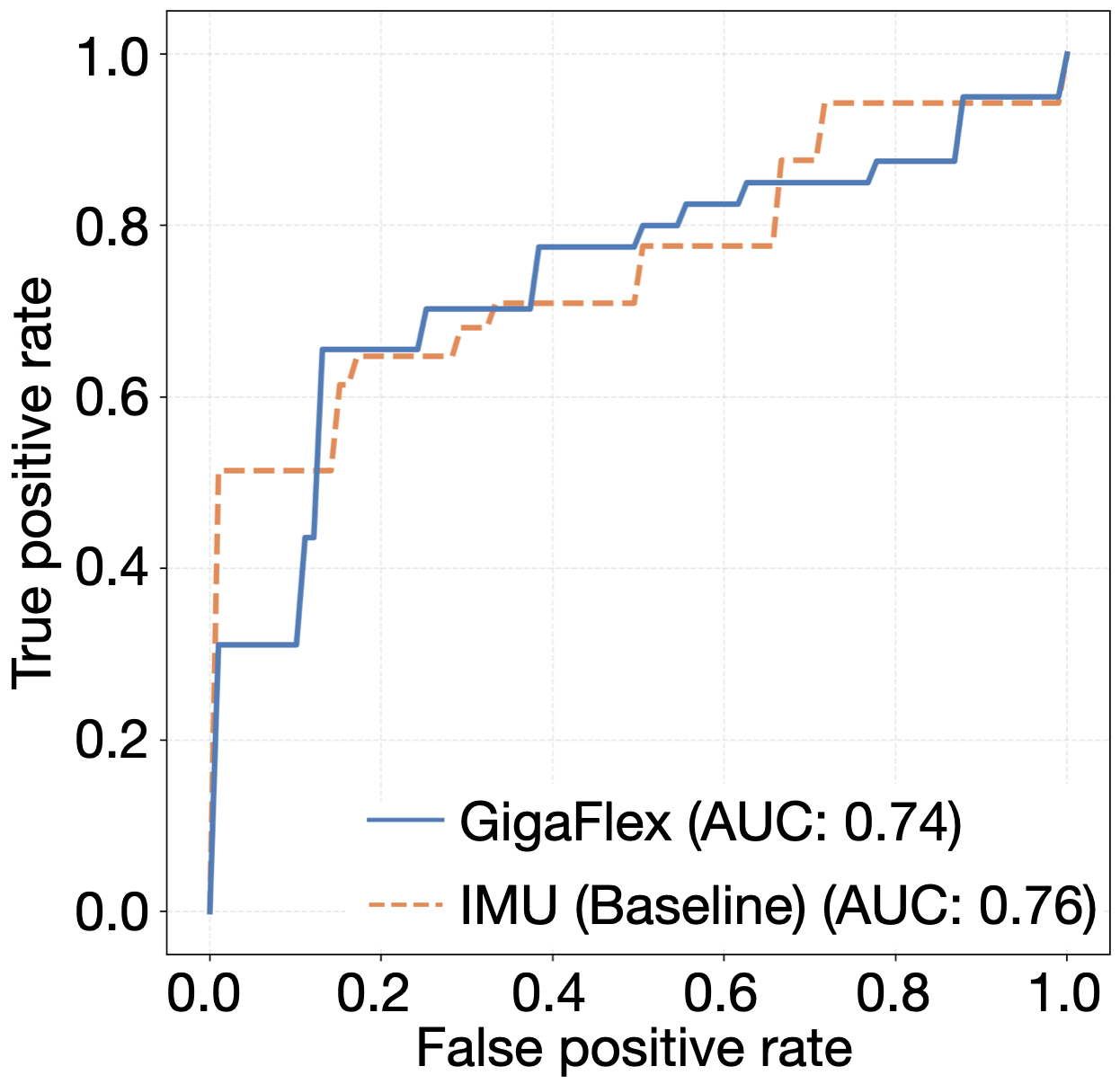}
    \vspace{-1em}
    \caption{{\bf Isotonic (dynamic) force evaluation.} We perform binary classification between light effort (10--30\% 1RM) and heavy effort (50--70\% 1RM).}
    \label{fig:isotonic}
\end{figure}

We evaluate isotonic force estimation accuracy using data collected during concentric bicep curl exercises, with loads determined by each subject's one-repetition maximum (1RM) (Fig.~\ref{fig:isotonic}). We perform binary force classification, and categorize 10\% and 30\% 1RM as \textit{light effort} and 50\% and 70\% 1RM as \textit{heavy effort}. We note that end-to-end force estimation during dynamic isotonic force estimation represents a challenging scenario due to substantial motion artifacts of the arm. We show in Fig.~\ref{fig:isotonic} that our contactless approach achieves an AUC of 0.74, which is comparable to the contact-based IMU baseline (AUC = 0.75), even in the presence of dynamic motion.




\subsection{Fatigue detection performance}
\label{sec:fatigue}

\begin{figure*}[ht]
    \centering
    \includegraphics[width=.7\textwidth]{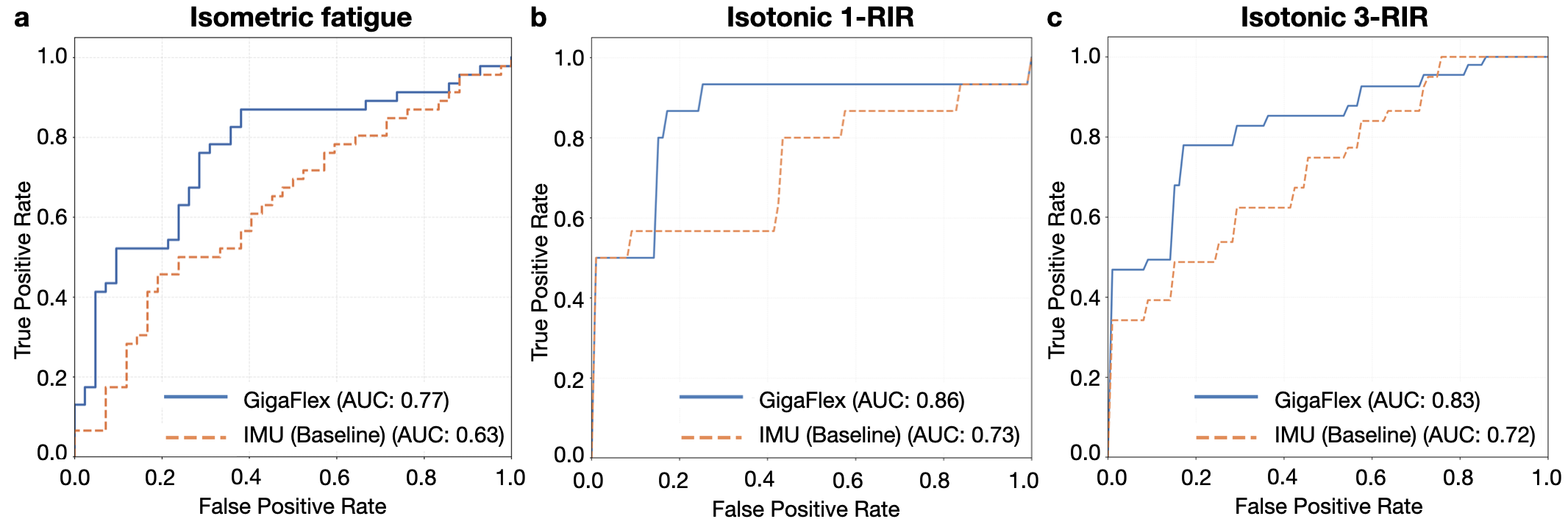}
    \vspace{-1em}
    \caption{{\bf Fatigue detection performance measured using receiver-operating curves for {\bf (a)} isometric fatigue {\bf (b) } isotonic 1-RIR and {\bf (c)} isotonic 3-RIR. RIR is the number of repetitions an individual can perform before muscular failure.}}
    \label{fig:fatigue}
\end{figure*}

\begin{figure*}[ht]
    \centering
    \includegraphics[width=\textwidth]{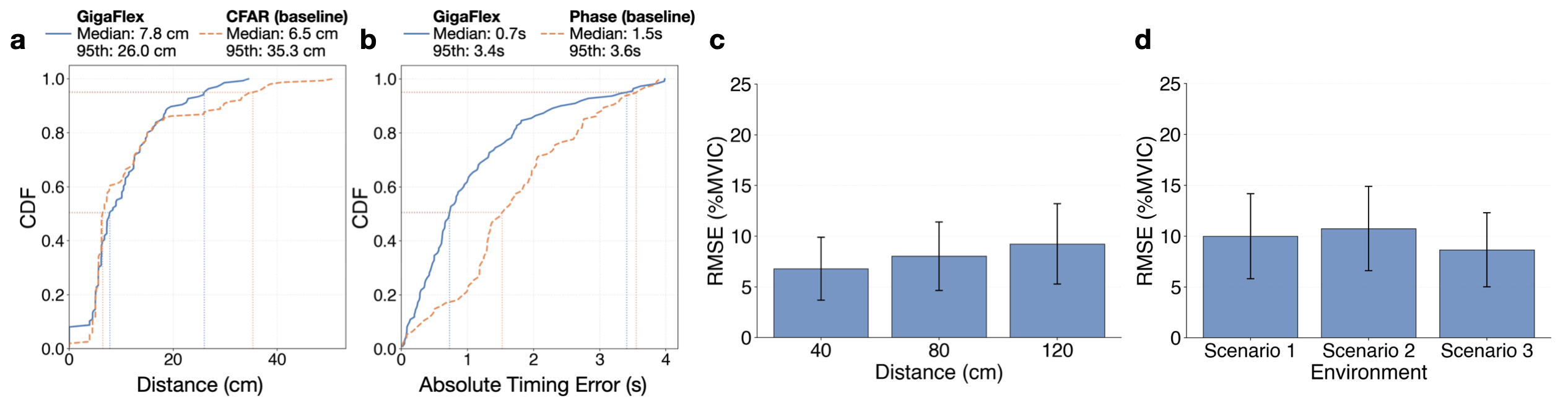}
    \vspace{-1em}
    \caption{{\bf Benchmarks.} {\bf (a,b)} Comparison of Chaos-informed vs. conventional localization and segmentation algorithms. {\bf (c,d)} Effect of distance and environment on MVIC error.}
    \label{fig:bm}
\end{figure*}



\subsubsection{Detecting fatigue from isometric exercise}

We profile each participant's muscle force outputs before and after inducing fatigue through isometric exercise. Fig.~\ref{fig:fatigue}a shows ROC curves comparing our radar-based system against contact-based IMU baseline measurements. Our approach achieves an AUC of 0.77, outperforming the wearable baseline (AUC = 0.63) by 22\%.

\subsubsection{Predicting proximity to muscle failure}


We evaluate our system's ability to detect proximity to muscle failure 1 and 3 repetitions before it occurs (also known as 1-RIR and 3-RIR respectively). 1-RIR corresponds to near maximal exertion and imminent muscle failure, while 3-RIR corresponds to high intensity effort and approaching muscle failure. We show in Fig.~\ref{fig:fatigue}b,c that using logistic regression with Chaos features from detrended fluctuation analysis (DFA), our system achieves 88\% accuracy (AUC = 0.86) for 1-RIR detection and 77\% (AUC = 0.83) for 3-RIR detection, compared to 82\% (AUC = 0.73) and 64\% (AUC = 0.72) respectively for IMU-based sensing. We note that the DFA features are valuable as they quantify long-range correlations in muscle activation patterns that deteriorate progressively with fatigue accumulation.




\subsection{Benchmarks}

\noindent {\bf Spatial localization performance.} We assess the performance of our entropic sieve for muscle localization during dynamic exercises. Fig.~\ref{fig:bm}a compares our two-stage pipeline combining CFAR and an entropy-based filter against the conventional approach of only using a CFAR-selected region with maximum signal strength. Our approach and the conventional approach achieve comparable median errors of 7.8 and 6.5~cm respectively, while the 95th percentile errors improves from 35.3 to 26.0~cm.

\noindent {\bf Temporal segmentation performance.} We evaluate the performance of our determinism-based (DET) approach for detecting repetition boundaries during resistance training. Our results in Fig.~\ref{fig:bm}b show a median temporal error of 0.7 s for our Chaos-informed radar method, which outperforms the radar phase signal baseline which has a 1.5 s error. When participants near muscular failure, their muscle vibrations become more irregular. However this only has a modest effect on our temporal segmentation, increasing error by only 0.164~s when compared to a participant's non-fatigued state.


\noindent\textbf{Effect of distance.} We evaluated MVIC error for a single subject at distances of 40, 80, and 120~cm from the radar for isometric elbow flexion.  We show in Fig.~\ref{fig:bm}c that average RMSE increases slightly with across the tested distances from approximately 7\% to 9\%. The results demonstrate that for practical operating distances, MVIC error are within acceptable use.

\noindent\textbf{Effect of different environments.} Fig.~\ref{fig:bm}d shows system performance across three distinct indoor environments. The modest variation in RMSE of approximately 2\% between scenarios show the system works well across different environmental conditions.

\section{Discussion, limitations, and future directions}

\noindent {\bf Evaluation on individuals with muscle injuries.} An important potential application of this technology lies in monitoring the progress of physical rehabilitation for patients with muscle injuries such as strains, bruising or anterior cruciate ligament (ACL) tears~\cite{alrushud_cross-sectional_2024}. In ACL rehabilitation in particular, even after meeting standard clinical discharge criteria, athletes often exhibit residual biomechanical asymmetries in muscle force generation that can be measured during vertical jumps, particularly during the concentric (push-off) phase~\cite{kotsifaki2023performance,king2019back,hewit2012asymmetry}. While such asymmetries are conventionally measured using large and expensive force plates~\cite{bertec,forcedecks}, the ability to track these asymmetries using our radar system could be a convenient way to track rehabilitation progress and reinjury risk particularly in a home setting~\cite{brunst_return--sport_2022, legnani_limb_2024, eagle_bilateral_2019}.

\noindent {\bf Scaling to complex exercise scenarios.} A limitation of our system is that it has only been evaluated on a single type of isometric and isotonic exercise, specifically elbow flexion and concentration bicep curls respectively, in controlled lab settings. The reason for this is that these exercises minimize confounding factors such as posture variation and occlusion, providing a controlled experimental setup suitable for developing a proof-of-concept radar system for tracking muscle force output. A natural extensions to this work would consider how it can adapt to exercises of increasing diversity and complexity, including: 

\squishlist
\item \textit{Across muscle groups.} \sysname\ currently measures muscle output at the arms where the vibration vectors of adjacent muscle groups are oriented in slightly different directions. Future work could extend this to other regions of the body including legs, abdomen and chest. This would require addressing new sources of cross-talk and interference. For instance, the effects of breathing and heartbeats would be more pronounced at the chest, while the abdomen has several closely spaced muscles facing in the same direction. 

\item \textit{Across exercise modalities.} Our system has not been evaluated on aerobic or full-body exercises involving large body movements such as running, cycling, or rowing. While there have been recent papers~\cite{bauder2025mm,gong2021rf} on mmWave sensing for respiration and heartbeat under significant motion artifacts such as running, future work is needed to investigate the extent to which mmWave sensing could be used for muscle sensing in high-mobility and motion-rich scenarios.

\item \textit{Across realworld environments.} Our system was evaluated in a controlled lab setting. Adapting the system so it can be deployed in gym, clinic or home environments is important to assess the system's robustness to diverse environmental geometries, equipment setups, and human motion in the vicinity. In particular, integrating the device as part of existing gym equipment (e.g. bicep curl machines, preacher curl benches, and cable curl stations) would enable real-world evaluation, under additional challenges such as reflection and scattering from metallic frames from the equipment.
\squishend

\noindent {\bf Flexible metasurfaces for multi-muscle detection.} Prior work on flexible metasurfaces~\cite{asyari202460}, which was originally developed to aid mmWave-based wrist pulse monitoring by amplifying the pulse motion, could be adapted for for muscle sensing and applied in a sticker or tattoo-like form factor. In particular, such metasurfaces could aid in simultaneously detecting the vibrations of multiple muscles, including those that may not be directly aligned with the radar.

\section{Related Work}

\begin{table}[h]
\centering
\small
\renewcommand{\arraystretch}{1.2}
\begin{tabularx}{0.45\textwidth}{|L|X|C|C|C|C|}
\hline
\textbf{Paper} & \textbf{Tech-nology} & \textbf{Contact-less} & \textbf{Iso-metric static \hspace{1em} force estimation} & \textbf{Isotonic dynamic force estimation} & \textbf{Fatigue detection} \\ \hline
\cite{velloso2013qualitative} & Cameras & \cmark & \xmark & \xmark & \xmark \\ \hline
\cite{lu2023investigation} & IMU + Ultrasound & \xmark & \cmark & \cmark & \xmark \\ \hline
\cite{khurana2018gymcam} & Cameras & \cmark & \xmark & \xmark & \xmark \\ \hline
\cite{song_myomonitor_2021} & Active sonar & \cmark & \xmark & \xmark & \cmark \\ \hline
\cite{tsengwa_toward_2024} & mmWave radar & \cmark & \xmark & \xmark & \xmark \\ \hline
\cite{wu2025bandei} & Electrical impedance tomography & \xmark & \xmark & \xmark & \xmark \\ \hline
\cite{mahmoodi_echoforce_2025} & Active sonar & \xmark & \cmark & \xmark & \xmark \\ \hline
\rowcolor{gray!15}
{\bf {\sysname} (ours)} & mmWave radar & \cmark & \cmark & \cmark & \cmark \\ \hline
\end{tabularx}
\caption{{\bf Comparison of representative muscle sensing systems.} {Uniquely, {\sysname} performs contactless force estimation for for elbow flexion (static) and bicep curls (dynamic), as well as fatigue detection.}}
\label{tab:comparison}
\vspace*{-0.2in}
\end{table}

\noindent {\bf Conventional Contact-Based Approaches.} Contact-based approaches broadly measure: (1) Electrical signals from muscle activation using Electromyography (EMG)~\cite{mills2005basics,de2006electromyography} including at high sampling rates~\cite{clancy2002sampling,kahl2015effects,durkin2005effects,li2010selection}. Intramuscular EMG requires needle insertion into the tissue and in general EMG requires direct skin contact with electrodes. (2) Mechanical vibration sensing using Mechanomyography (MMG) through accelerometers or microphones~\cite{islam2013mechanomyogram,silva2004mmg}. In contrast, \sysname\ measures muscle vibrations in an entriely contact-free manner that is significantly less invasive.

\noindent {\bf Vision-Based Systems.} Camera-based approaches estimate muscle engagement from body pose and movement patterns captured by smartphone or depth cameras~\cite{sharshar2023camera,khurana2018gymcam,ganesh2020personalized,zhang2011physical}. However, unlike \sysname, vision-based methods perform only skeletal tracking and can only indirectly infer muscle force output or fatigue onset.

\noindent {\bf Wearable Systems.} Wearable IMU and acoustic sensors embedded in smartwatches and fitness bands infer muscle exertion through motion analysis and activity classification~\cite{mahmoodi_echoforce_2025,lu2023investigation,velloso2013qualitative}. Smart garments integrate flexible, textile-based sensors leveraging strain gauges or electrical impedance sensing to detect muscle deformation during exercise~\cite{belbasis2015muscle,wang2015smart,wu2025bandei,alvarez2022towards,almohimeed2013development}. Despite their portability, wearable sensors can be uncomfortable to wear for long periods of time.

\noindent {\bf Contactless Muscle Sensing.} Prior contact-free muscle tissue systems include: (1) Laser Doppler vibrometry (LDV) can remotely capture skin displacements caused by cardiac or muscular activity~\cite{Sara_LDV}. However, LDV systems require precise optical alignment and are highly sensitive to occlusions, making them impractical under natural or dynamic conditions. (2) Song et al.~\cite{song_myomonitor_2021} uses active sonar on smartphones to detect muscle fatigue, but requires positioning the smartphone close to the target muscle (5--10~cm) with the display directly facing the skin, limiting its practicality for use in everyday environments {and multi-muscle monitoring.}

\noindent {\bf RF and Radar-based Sensing.} Radio-frequency (RF)–based vibration sensing, particularly with millimeter-wave (mmWave) radar -- with its millimeter-scale wavelength -- can perform vibration sensing at micrometer scale over a wide field of view, and has been applied in vital-sign monitoring using and has demonstrated the ability to detect breathing~\cite{zhicheng_mmWave_vitalsign}, heartbeat~\cite{Langcheng_mmArrhythmia, fadel_smartHome}, and even pulse-induced displacements at sites such as the neck and wrist with micrometer-level precision~\cite{geng_CaPTT_2023, liang_airbp_2023, zhu2025measuring}. However, these works focus on extracting periodic features in a narrow-band under 5~Hz range. In contrast, muscle vibrations are broadband (1--100~Hz), non-periodic, and exhibit highly dynamic spatial-temporal patterns.


More recently, mmWave radar has been explored for muscle-related sensing tasks. The signal amplitude has been used to classify Parkinsonian versus essential tremor~\cite{gillani2023Tumor}, while radar phase information has been compared with surface electromyography (sEMG) to estimate muscle activation~\cite{tsengwa_toward_2024}. The latter represents an important early step toward radar-based muscle monitoring; however, these approaches primarily focus on coarse muscle activation detection rather than characterizing the continuous mechanical dynamics underlying contraction. In contrast, by leveraging nonlinear dynamic analysis and chaos-theoretic modeling, we enable estimation of muscle force output and fatigue detection directly from radar reflections, without any on-body instrumentation.



\section{Conclusion }
This paper presents \sysname, a contactless system to monitor muscle vibrations during exercise. \sysname\ infuses principles from Chaos theory and non-linear dynamics to process radar signals temporally and spatially to extract quantitative metrics on muscle fatigue. A detailed implementation and user study demonstrates low errors in key muscle fatigue metrics, comparable to state-of-the-art contact-based sensing systems. As next steps, we seek to: (1) Evaluate our system on individuals with muscle injuries; (2) Scale our system to diverse exercise scenarios where multiple muscle groups are activated; (3) Accelerate system performance and latency for real-time operation.


\section*{References}
\begin{thebibliography}{100}

\bibitem{lourenco_relationship_2023}
João Lourenço, Élvio~Rúbio Gouveia, Hugo Sarmento, Andreas Ihle, Tiago Ribeiro, Ricardo Henriques, Francisco Martins, Cíntia França, Ricardo~Maia Ferreira, Luís Fernandes, Pedro Teques, and Daniel Duarte.
\newblock Relationship between {Objective} and {Subjective} {Fatigue} {Monitoring} {Tests} in {Professional} {Soccer}.
\newblock {\em International Journal of Environmental Research and Public Health}, 20(2):1539, January 2023.
\newblock Publisher: Multidisciplinary Digital Publishing Institute.

\bibitem{Loy_Relatio_2017}
Bryan Loy, Ruby Taylor, Brett Fling, and Fay Horak.
\newblock Relationship between perceived fatigue and performance fatigability in people with multiple sclerosis: A systematic review and meta-analysis.
\newblock {\em Journal of Psychosomatic Research}, 100, 06 2017.

\bibitem{islam_prevalence_2024}
Mohammad~Jahirul Islam, Md~Selim Rana, Md~Sharifuddin Sarker, Md~Mahemanul Islam, Md~Nuruzzaman Miah, Md~Anwar Hossain, Ruwaida Jahangir, Rahemun Akter, and Sohel Ahmed.
\newblock Prevalence and predictors of musculoskeletal injuries among gym members in {Bangladesh}: {A} nationwide cross-sectional study.
\newblock {\em PLOS ONE}, 19(8):e0303461, August 2024.
\newblock Publisher: Public Library of Science.

\bibitem{lubetzky2009prevalence}
A~Lubetzky-Vilnai, E~Carmeli, and Michal Katz-Leurer.
\newblock Prevalence of injuries among young adults in sport centers: relation to the type and pattern of activity.
\newblock {\em Scandinavian journal of medicine \& science in sports}, 19(6):828--833, 2009.

\bibitem{sharma_prevalence_2024}
Tamana Sharma and Keerthi Rao.
\newblock {PREVALENCE} {OF} {MUSCULOSKELETAL} {INJURIES} {IN} {GYM} {GOING} {ADULTS}: {A} {SURVEY} {STUDY}.
\newblock 1, May 2024.
\newblock Publisher: Mendeley Data.

\bibitem{alrushud_cross-sectional_2024}
Asma~Saad Alrushud.
\newblock A cross-sectional study of musculoskeletal injuries related to exercise among gym members in {Saudi} {Arabia} in 2022: prevalence, common types, and predictor factors.
\newblock {\em BMC Musculoskeletal Disorders}, 25(1):621, August 2024.

\bibitem{sayyadi_effectiveness_2024}
Parisa Sayyadi, Hooman Minoonejad, Foad Seidi, Rahman Shikhhoseini, and Ramin Arghadeh.
\newblock The effectiveness of fatigue on repositioning sense of lower extremities: systematic review and meta-analysis.
\newblock {\em BMC Sports Science, Medicine and Rehabilitation}, 16(1):35, February 2024.

\bibitem{hermens2000development}
Hermie~J Hermens, Bart Freriks, Catherine Disselhorst-Klug, and G{\"u}nter Rau.
\newblock Development of recommendations for semg sensors and sensor placement procedures.
\newblock {\em Journal of electromyography and Kinesiology}, 10(5):361--374, 2000.

\bibitem{daube2009needle}
Jasper~R Daube and Devon~I Rubin.
\newblock Needle electromyography.
\newblock {\em Muscle \& Nerve: Official Journal of the American Association of Electrodiagnostic Medicine}, 39(2):244--270, 2009.

\bibitem{hallock2021toward}
Laura~A Hallock, Bhavna Sud, Chris Mitchell, Eric Hu, Fayyaz Ahamed, Akash Velu, Amanda Schwartz, and Ruzena Bajcsy.
\newblock Toward real-time muscle force inference and device control via optical-flow-tracked muscle deformation.
\newblock {\em IEEE Transactions on Neural Systems and Rehabilitation Engineering}, 29:2625--2634, 2021.

\bibitem{shi2008continuous}
Jun Shi, Yong-Ping Zheng, Qing-Hua Huang, and Xin Chen.
\newblock Continuous monitoring of sonomyography, electromyography and torque generated by normal upper arm muscles during isometric contraction: sonomyography assessment for arm muscles.
\newblock {\em IEEE transactions on biomedical engineering}, 55(3):1191--1198, 2008.

\bibitem{kamatham2022simple}
Anne~Tryphosa Kamatham, Meena Alzamani, Allison Dockum, Siddhartha Sikdar, and Biswarup Mukherjee.
\newblock A simple, drift compensated method for estimation of isometric force using sonomyography.
\newblock In {\em Sensing Technology: Proceedings of ICST 2022}, pages 355--366. Springer, 2022.

\bibitem{hallock2020muscle}
Laura~A Hallock, Akash Velu, Amanda Schwartz, and Ruzena Bajcsy.
\newblock Muscle deformation correlates with output force during isometric contraction.
\newblock In {\em 2020 8th IEEE RAS/EMBS International Conference for Biomedical Robotics and Biomechatronics (BioRob)}, pages 1188--1195. IEEE, 2020.

\bibitem{brausch2022classifying}
Lukas Brausch, Holger Hewener, and Paul Lukowicz.
\newblock Classifying muscle states with one-dimensional radio-frequency signals from single element ultrasound transducers.
\newblock {\em Sensors}, 22(7):2789, 2022.

\bibitem{woodward2019segmenting}
Richard~B Woodward, Maria~J Stokes, Sandra~J Shefelbine, and Ravi Vaidyanathan.
\newblock Segmenting mechanomyography measures of muscle activity phases using inertial data.
\newblock {\em Scientific reports}, 9(1):5569, 2019.

\bibitem{barry1985acoustic}
Daniel~T Barry, Steven~R Geiringer, and Richard~D Ball.
\newblock Acoustic myography: a noninvasive monitor of motor unit fatigue.
\newblock {\em Muscle \& Nerve: Official Journal of the American Association of Electrodiagnostic Medicine}, 8(3):189--194, 1985.

\bibitem{Sara_LDV}
Sara Casaccia, Lorenzo Scalise, Luigi Casacanditella, Enrico~P. Tomasini, and John~W. Rohrbaugh.
\newblock Non-contact assessment of muscle contraction: Laser doppler myography.
\newblock In {\em 2015 IEEE International Symposium on Medical Measurements and Applications (MeMeA) Proceedings}, pages 610--615, 2015.

\bibitem{song_myomonitor_2021}
Xingzhe Song, Hongshuai Li, and Wei Gao.
\newblock {MyoMonitor}: {Evaluating} muscle fatigue with commodity smartphones.
\newblock {\em Smart Health}, 19:100175, March 2021.

\bibitem{Langcheng_mmArrhythmia}
Langcheng Zhao, Rui Lyu, Qi~Lin, Anfu Zhou, Huanhuan Zhang, Huadong Ma, Jingjia Wang, Chunli Shao, and Yida Tang.
\newblock mmarrhythmia: Contactless arrhythmia detection via mmwave sensing.
\newblock {\em Proc. ACM Interact. Mob. Wearable Ubiquitous Technol.}, 8(1), March 2024.

\bibitem{fadel_smartHome}
Fadel Adib, Hongzi Mao, Zachary Kabelac, Dina Katabi, and Robert~C. Miller.
\newblock Smart homes that monitor breathing and heart rate.
\newblock In {\em Proceedings of the 33rd Annual ACM Conference on Human Factors in Computing Systems}, CHI '15, page 837–846, New York, NY, USA, 2015. Association for Computing Machinery.

\bibitem{geng_CaPTT_2023}
Fanglin Geng, Zhongrui Bai, Hao Zhang, Yicheng Yao, Changyu Liu, Peng Wang, Xianxiang Chen, Lidong Du, Xiaoran Li, Baoshi Han, and Zhen Fang.
\newblock Contactless and continuous blood pressure measurement according to {caPTT} obtained from millimeter wave radar.
\newblock {\em Measurement}, 218:113151, August 2023.

\bibitem{liang_airbp_2023}
Yumeng Liang, Anfu Zhou, Xinzhe Wen, Wei Huang, Pu~Shi, Lingyu Pu, Huanhuan Zhang, and Huadong Ma.
\newblock {airBP}: {Monitor} {Your} {Blood} {Pressure} with {Millimeter}-{Wave} in the {Air}.
\newblock {\em ACM Transactions on Internet of Things}, 4(4):28:1--28:32, November 2023.

\bibitem{zhu2025measuring}
Jiangyifei Zhu, Kuang Yuan, Akarsh Prabhakara, Yunzhi Li, Gongwei Wang, Kelly Michaelsen, Justin Chan, and Swarun Kumar.
\newblock Measuring multi-site pulse transit time with an ai-enabled mmwave radar.
\newblock {\em arXiv preprint arXiv:2510.18141}, 2025.

\bibitem{zhao2018noncontact}
Heng Zhao, Hong Hong, Dongyu Miao, Yusheng Li, Haitao Zhang, Yingming Zhang, Changzhi Li, and Xiaohua Zhu.
\newblock A noncontact breathing disorder recognition system using 2.4-ghz digital-if doppler radar.
\newblock {\em IEEE journal of biomedical and health informatics}, 23(1):208--217, 2018.

\bibitem{li2022detection}
Gen Li, Yun Ge, Yiyu Wang, Qingwu Chen, and Gang Wang.
\newblock Detection of human breathing in non-line-of-sight region by using mmwave fmcw radar.
\newblock {\em IEEE Transactions on Instrumentation and Measurement}, 71:1--11, 2022.

\bibitem{tsengwa_toward_2024}
Kukhokuhle Tsengwa, Stephen Paine, Fred Nicolls, Yumna Albertus, and Amir Patel.
\newblock Toward {Noncontact} {Muscle} {Activity} {Estimation} {Using} {FMCW} {Radar}.
\newblock {\em IEEE Sensors Journal}, 24(22), November 2024.

\bibitem{gitter1995fractal}
J~Andrew Gitter and M~Joseph Czerniecki.
\newblock Fractal analysis of the electromyographic interference pattern.
\newblock {\em Journal of neuroscience Methods}, 58(1-2):103--108, 1995.

\bibitem{nieminen1996evidence}
H~Nieminen and EP~Takala.
\newblock Evidence of deterministic chaos in the myoelectric signal.
\newblock {\em Electromyography and clinical neurophysiology}, 36(1):49--58, 1996.

\bibitem{enoka_muscle_2008}
Roger~M. Enoka and Jacques Duchateau.
\newblock Muscle fatigue: what, why and how it influences muscle function.
\newblock {\em The Journal of Physiology}, 586(1):11--23, 2008.
\newblock \_eprint: https://physoc.onlinelibrary.wiley.com/doi/pdf/10.1113/jphysiol.2007.139477.

\bibitem{khodadadi2023nonlinear}
Vahid Khodadadi, Fereidoun~Nowshiravan Rahatabad, Ali Sheikhani, and Nader~Jafarnia Dabanloo.
\newblock Nonlinear analysis of biceps surface emg signals for chaotic approaches.
\newblock {\em Chaos, Solitons \& Fractals}, 166:112965, 2023.

\bibitem{xie2009detection}
Hong-Bo Xie, Yong-Ping Zheng, and Guo Jing-Yi.
\newblock Detection of chaos in human fatigue mechanomyogarphy signals.
\newblock In {\em 2009 Annual International Conference of the IEEE Engineering in Medicine and Biology Society}, pages 4379--4382. IEEE, 2009.

\bibitem{filligoi2011chaos}
G~Filligoi.
\newblock Chaos theory and semg.
\newblock {\em JSAB January Edition}, 2011.

\bibitem{xiong2014application}
Anbin Xiong, Xingang Zhao, Jianda Han, and Guangjun Liu.
\newblock Application of the chaos theory in the analysis of emg on patients with facial paralysis.
\newblock In {\em Robot Intelligence Technology and Applications 2: Results from the 2nd International Conference on Robot Intelligence Technology and Applications}, pages 805--819. Springer, 2014.

\bibitem{conte2015chaos}
Elio Conte, Ken Ware, Riccardo Marvulli, Giancarlo Ianieri, Marisa Megna, Sergio Conte, Leonardo Mendolicchio, Enrico Pierangeli, et~al.
\newblock Chaos, fractal and recurrence quantification analysis of surface electromyography in muscular dystrophy.
\newblock {\em World Journal of Neuroscience}, 5(4):205--257, 2015.

\bibitem{lei2012nonlinear}
Min Lei and Guang Meng.
\newblock Nonlinear analysis of surface emg signals.
\newblock {\em Computational Intelligence in Electromyography Analysis—A Perspective on Current Applications and Future Challenges}, pages 120--171, 2012.

\bibitem{chen2007characterization}
Weiting Chen, Zhizhong Wang, Hongbo Xie, and Wangxin Yu.
\newblock Characterization of surface emg signal based on fuzzy entropy.
\newblock {\em IEEE Transactions on neural systems and rehabilitation engineering}, 15(2):266--272, 2007.

\bibitem{athos}
How to calibrate your athos apparel, 2025.

\bibitem{myo}
Tips and tricks for the myo armband, 2025.

\bibitem{islam2013mechanomyogram}
Md~Anamul Islam, Kenneth Sundaraj, R~Badlishah Ahmad, and Nizam~Uddin Ahamed.
\newblock Mechanomyogram for muscle function assessment: a review.
\newblock {\em PloS one}, 8(3):e58902, 2013.

\bibitem{matsumoto2025muscle}
Tatsuhiko Matsumoto, Yutaka Takamaru, Chikao Nakayama, Shuji Sawada, and Shuichi Machida.
\newblock Muscle fiber recruitment characteristics in trained older adults: An emg frequency analysis during voluntary contraction.
\newblock {\em Journal of Functional Morphology and Kinesiology}, 10(3):355, 2025.

\bibitem{orizio1996surface}
Claudio Orizio, Diego Liberati, Cecilia Locatelli, Domenico De~Grandis, and Arsenio Veicsteinas.
\newblock Surface mechanomyogram reflects muscle fibres twitches summation.
\newblock {\em Journal of biomechanics}, 29(4):475--481, 1996.

\bibitem{lubel2023non}
Emma Lubel, Bruno~Grandi Sgambato, Robin Rohl{\'e}n, Jaime Ib{\'a}{\~n}ez, Deren~Y Barsakcioglu, Meng-Xing Tang, and Dario Farina.
\newblock Non-linearity in motor unit velocity twitch dynamics: Implications for ultrafast ultrasound source separation.
\newblock {\em IEEE Transactions on Neural Systems and Rehabilitation Engineering}, 31:3699--3710, 2023.

\bibitem{beck2005mechanomyographic}
Travis~W Beck, Terry~J Housh, Joel~T Cramer, Joseph~P Weir, Glen~O Johnson, Jared~W Coburn, Moh~H Malek, and Michelle Mielke.
\newblock Mechanomyographic amplitude and frequency responses during dynamic muscle actions: a comprehensive review.
\newblock {\em Biomedical engineering online}, 4(1):67, 2005.

\bibitem{williams2017borg}
Nerys Williams.
\newblock The borg rating of perceived exertion (rpe) scale.
\newblock {\em Occupational medicine}, 67(5):404--405, 2017.

\bibitem{ritchie2012rating}
Carrie Ritchie.
\newblock Rating of perceived exertion (rpe).
\newblock {\em Journal of physiotherapy}, 58(1):62, 2012.

\bibitem{marcora2009mental}
Samuele~M Marcora, Walter Staiano, and Victoria Manning.
\newblock Mental fatigue impairs physical performance in humans.
\newblock {\em Journal of applied physiology}, 106(3):857--864, 2009.

\bibitem{smirmaul2012sense}
Bruno de Paula~Cara{\c{c}}a Smirmaul.
\newblock Sense of effort and other unpleasant sensations during exercise: clarifying concepts and mechanisms.
\newblock {\em British journal of sports medicine}, 46(5):308--311, 2012.

\bibitem{halperin2020rating}
Israel Halperin and Aviv Emanuel.
\newblock Rating of perceived effort: methodological concerns and future directions.
\newblock {\em Sports Medicine}, 50(4):679--687, 2020.

\bibitem{andersen2010muscle}
Lars~L Andersen, Christoffer~H Andersen, Ole~S Mortensen, Otto~M Poulsen, Inger Birthe~T Bj{\o}rnlund, and Mette~K Zebis.
\newblock Muscle activation and perceived loading during rehabilitation exercises: comparison of dumbbells and elastic resistance.
\newblock {\em Physical therapy}, 90(4):538--549, 2010.

\bibitem{isometric}
Guide to the top 20 isometric exercises for static strength training, 2025.

\bibitem{meldrum2007maximum}
Dara Meldrum, Eibhlis Cahalane, Ronan Conroy, Deirdre Fitzgerald, and Orla Hardiman.
\newblock Maximum voluntary isometric contraction: reference values and clinical application.
\newblock {\em Amyotrophic Lateral Sclerosis}, 8(1):47--55, 2007.

\bibitem{hunter2002task}
Sandra~K Hunter, Daphne~L Ryan, Justus~D Ortega, and Roger~M Enoka.
\newblock Task differences with the same load torque alter the endurance time of submaximal fatiguing contractions in humans.
\newblock {\em Journal of neurophysiology}, 88(6):3087--3096, 2002.

\bibitem{seo2012reliability}
Dong-il Seo, Eonho Kim, Christopher~A Fahs, Lindy Rossow, Kaelin Young, Steven~L Ferguson, Robert Thiebaud, Vanessa~D Sherk, Jeremy~P Loenneke, Daeyeol Kim, et~al.
\newblock Reliability of the one-repetition maximum test based on muscle group and gender.
\newblock {\em Journal of sports science \& medicine}, 11(2):221, 2012.

\bibitem{grgic2020test}
Jozo Grgic, Bruno Lazinica, Brad~J Schoenfeld, and Zeljko Pedisic.
\newblock Test--retest reliability of the one-repetition maximum (1rm) strength assessment: a systematic review.
\newblock {\em Sports medicine-open}, 6(1):1--16, 2020.

\bibitem{niewiadomski2008determination}
Wiktor Niewiadomski, Dorota Laskowska, Anna G{\k{a}}siorowska, Gerard Cybulski, Anna Strasz, and J{\'o}zef Langfort.
\newblock Determination and prediction of one repetition maximum (1rm): safety considerations.
\newblock {\em J Hum Kinet}, 19(2008):109--120, 2008.

\bibitem{roman2016influence}
Danuta Roman-Liu.
\newblock The influence of confounding factors on the relationship between muscle contraction level and mf and mpf values of emg signal: a review.
\newblock {\em International Journal of Occupational Safety and Ergonomics}, 22(1):77--91, 2016.

\bibitem{anders2019inter}
John Paul~V Anders, Cory~M Smith, Joshua~L Keller, Ethan~C Hill, Terry~J Housh, Richard~J Schmidt, and Glen~O Johnson.
\newblock Inter-and intra-individual differences in emg and mmg during maximal, bilateral, dynamic leg extensions.
\newblock {\em Sports}, 7(7):175, 2019.

\bibitem{rodrick2006nonlinear}
David Rodrick and Waldemar Karwowski.
\newblock Nonlinear dynamical behavior of surface electromyographical signals of biceps muscle under two simulated static work postures.
\newblock {\em Nonlinear dynamics, psychology, and life sciences}, 10(1):21--35, 2006.

\bibitem{padmanabhan2004nonlinear}
Pavitra Padmanabhan and Sadasivan Puthusserypady.
\newblock Nonlinear analysis of emg signals-a chaotic approach.
\newblock In {\em The 26th Annual International Conference of the IEEE Engineering in Medicine and Biology Society}, volume~1, pages 608--611. IEEE, 2004.

\bibitem{rampichini2020complexity}
Susanna Rampichini, Taian~Martins Vieira, Paolo Castiglioni, and Giampiero Merati.
\newblock Complexity analysis of surface electromyography for assessing the myoelectric manifestation of muscle fatigue: A review.
\newblock {\em Entropy}, 22(5):529, 2020.

\bibitem{schreiber1996improved}
Thomas Schreiber and Andreas Schmitz.
\newblock Improved surrogate data for nonlinearity tests.
\newblock {\em Physical review letters}, 77(4):635, 1996.

\bibitem{tucker1999lorenz}
Warwick Tucker.
\newblock The lorenz attractor exists.
\newblock {\em Comptes Rendus de l'Acad{\'e}mie des Sciences-Series I-Mathematics}, 328(12):1197--1202, 1999.

\bibitem{richman2004sample}
Joshua~S Richman, Douglas~E Lake, and J~Randall Moorman.
\newblock Sample entropy.
\newblock In {\em Methods in enzymology}, volume 384, pages 172--184. Elsevier, 2004.

\bibitem{grassberger1983measuring}
Peter Grassberger and Itamar Procaccia.
\newblock Measuring the strangeness of strange attractors.
\newblock {\em Physica D: nonlinear phenomena}, 9(1-2):189--208, 1983.

\bibitem{marwan2007recurrence}
Norbert Marwan, M~Carmen Romano, Marco Thiel, and J{\"u}rgen Kurths.
\newblock Recurrence plots for the analysis of complex systems.
\newblock {\em Physics reports}, 438(5-6):237--329, 2007.

\bibitem{alizadeh_mmWaveBodyPenetration_2019}
Mostafa Alizadeh, George Shaker, Joao Carlos Martins~De Almeida, Plinio~Pelegrini Morita, and Safeddin Safavi-Naeini.
\newblock Remote {Monitoring} of {Human} {Vital} {Signs} {Using} mm-{Wave} {FMCW} {Radar}.
\newblock {\em IEEE Access}, 7:54958--54968, 2019.

\bibitem{zhao_emotion_2016}
Mingmin Zhao, Fadel Adib, and Dina Katabi.
\newblock Emotion recognition using wireless signals.
\newblock In {\em Proceedings of the 22nd {Annual} {International} {Conference} on {Mobile} {Computing} and {Networking}}, pages 95--108, New York City New York, October 2016. ACM.

\bibitem{diff}
Noise robust differentiators for second derivative estimation., 2025.

\bibitem{wendi2018extended}
Dadiyorto Wendi and Norbert Marwan.
\newblock Extended recurrence plot and quantification for noisy continuous dynamical systems.
\newblock {\em Chaos: An Interdisciplinary Journal of Nonlinear Science}, 28(8):085722, 2018.

\bibitem{richards2014radar}
Mark~A. Richards.
\newblock {\em Fundamentals of Radar Signal Processing}.
\newblock McGraw-Hill Education, New York, 2nd edition, 2014.

\bibitem{xu2022vitpose}
Yufei Xu, Jing Zhang, Qiming Zhang, and Dacheng Tao.
\newblock Vi{TP}ose: Simple vision transformer baselines for human pose estimation.
\newblock In {\em Advances in Neural Information Processing Systems}, 2022.

\bibitem{naschitz2004patterns}
JE~Naschitz, I~Rosner, M~Rozenbaum, M~Fields, H~Isseroff, JP~Babich, E~Zuckerman, N~Elias, D~Yeshurun, S~Naschitz, et~al.
\newblock Patterns of cardiovascular reactivity in disease diagnosis.
\newblock {\em Qjm}, 97(3):141--151, 2004.

\bibitem{marwan2002recurrence}
Norbert Marwan, Niels Wessel, Udo Meyerfeldt, Alexander Schirdewan, and J{\"u}rgen Kurths.
\newblock Recurrence-plot-based measures of complexity and their application to heart-rate-variability data.
\newblock {\em Physical review E}, 66(2):026702, 2002.

\bibitem{webber1994dynamical}
Charles~L Webber~Jr and Joseph~P Zbilut.
\newblock Dynamical assessment of physiological systems and states using recurrence plot strategies.
\newblock {\em Journal of applied physiology}, 76(2):965--973, 1994.

\bibitem{hu2018squeeze}
Jie Hu, Li~Shen, and Gang Sun.
\newblock Squeeze-and-excitation networks.
\newblock In {\em Proceedings of the IEEE conference on computer vision and pattern recognition}, pages 7132--7141, 2018.

\bibitem{Walton2002}
DM~Walton, JM~Elliott, GZ~Heller, M~Lee, and M~Sterling.
\newblock Pressure pain thresholds, force, and pain during isometric muscle contraction.
\newblock {\em Journal of Pain}, 3(5):324--332, 2002.

\bibitem{Refalo2023}
Martin~C. Refalo, Eric~R. Helms, D.~Lee Hamilton, Jackson~J. Fyfe, Daniel~W. Hart, David Bishop, Rafael Garcia‐Ros, Katrin Doma, Lucas~A. Gallo, Javier~P. Wagle, et~al.
\newblock Influence of resistance training proximity‐to‐failure, determined by repetitions‐in‐reserve, on neuromuscular fatigue in resistance‐trained males and females.
\newblock {\em Sports Medicine - Open}, 9(1):10, 2023.

\bibitem{ti_radar}
Awr1843boost, 2025.

\bibitem{adc}
Ads1256, 2025.

\bibitem{arduino_r4}
Arduino uno r4 minima, 2025.

\bibitem{imu}
6-axis imu (inertial measurement unit) with embedded ai: always-on 3-axis accelerometer and 3-axis gyroscope, 2025.

\bibitem{due}
Arduino due, 2025.

\bibitem{arducam}
Awr1843boost, 2025.

\bibitem{bmi}
Adult bmi categories, 2025.

\bibitem{frisancho1990anthropometric}
A~Roberto Frisancho.
\newblock {\em Anthropometric standards for the assessment of growth and nutritional status}.
\newblock University of Michigan press, 1990.

\bibitem{heymsfield1982anthropometric}
Steven Heymsfield, Cliford McManus, Janet Smith, Victoria Stevens, and Daniel~W Nixon.
\newblock Anthropometric measurement of muscle mass: revised equations for calculating bone-free arm muscle area.
\newblock {\em The American journal of clinical nutrition}, 36(4):680--690, 1982.

\bibitem{orizio_surface_2003}
Claudio Orizio, Massimiliano Gobbo, Bertrand Diemont, Fabio Esposito, and Arsenio Veicsteinas.
\newblock The surface mechanomyogram as a tool to describe the influence of fatigue on biceps brachii motor unit activation strategy. {Historical} basis and novel evidence.
\newblock {\em European Journal of Applied Physiology}, 90(3):326--336, October 2003.

\bibitem{androulakis_korakakis_optimizing_2024}
Patroklos Androulakis~Korakakis, Milo Wolf, Max Coleman, Ryan Burke, Alec Piñero, Jeff Nippard, and Brad~J. Schoenfeld.
\newblock Optimizing {Resistance} {Training} {Technique} to {Maximize} {Muscle} {Hypertrophy}: {A} {Narrative} {Review}.
\newblock {\em Journal of Functional Morphology and Kinesiology}, 9(1):9, March 2024.
\newblock Publisher: Multidisciplinary Digital Publishing Institute.

\bibitem{allen_skeletal_2008}
D.~G. Allen, G.~D. Lamb, and H.~Westerblad.
\newblock Skeletal {Muscle} {Fatigue}: {Cellular} {Mechanisms}.
\newblock {\em Physiological Reviews}, 88(1):287--332, January 2008.

\bibitem{kotsifaki2023performance}
Roula Kotsifaki, Vasileios Sideris, Enda King, Roald Bahr, and Rod Whiteley.
\newblock Performance and symmetry measures during vertical jump testing at return to sport after acl reconstruction.
\newblock {\em British Journal of Sports Medicine}, 57(20):1304--1310, 2023.

\bibitem{king2019back}
Enda King, Chris Richter, Andy Franklyn-Miller, Ross Wadey, Ray Moran, and Siobhan Strike.
\newblock Back to normal symmetry? biomechanical variables remain more asymmetrical than normal during jump and change-of-direction testing 9 months after anterior cruciate ligament reconstruction.
\newblock {\em The American journal of sports medicine}, 47(5):1175--1185, 2019.

\bibitem{hewit2012asymmetry}
Jennifer~K Hewit, John~B Cronin, and Patria~A Hume.
\newblock Asymmetry in multi-directional jumping tasks.
\newblock {\em Physical Therapy in Sport}, 13(4):238--242, 2012.

\bibitem{bertec}
Force plates, 2025.

\bibitem{forcedecks}
Forcedecks, 2025.

\bibitem{brunst_return--sport_2022}
Caroline Brunst, Matthew~P. Ithurburn, Andrew~M. Zbojniewicz, Mark~V. Paterno, and Laura~C. Schmitt.
\newblock Return-to-sport quadriceps strength symmetry impacts 5-year cartilage integrity after anterior cruciate ligament reconstruction: {A} preliminary analysis.
\newblock {\em Journal of Orthopaedic Research}, 40(1):285--294, 2022.
\newblock \_eprint: https://onlinelibrary.wiley.com/doi/pdf/10.1002/jor.25029.

\bibitem{legnani_limb_2024}
Claudio Legnani, Matteo Del~Re, Giuseppe~M. Peretti, Enrico Borgo, Vittorio Macchi, and Alberto Ventura.
\newblock Limb asymmetries persist 6 months after anterior cruciate ligament reconstruction according to the results of a jump test battery.
\newblock {\em Frontiers in Medicine}, 11, February 2024.
\newblock Publisher: Frontiers.

\bibitem{eagle_bilateral_2019}
Shawn~R. Eagle, Karen~A. Keenan, Chris Connaboy, Meleesa Wohleber, Andrew Simonson, and Bradley~C. Nindl.
\newblock Bilateral {Quadriceps} {Strength} {Asymmetry} {Is} {Associated} {With} {Previous} {Knee} {Injury} in {Military} {Special} {Tactics} {Operators}.
\newblock {\em The Journal of Strength \& Conditioning Research}, 33(1):89, January 2019.

\bibitem{bauder2025mm}
Chandler~Jackson Bauder, Tianhao Wu, Syed Irfan~Ali Meerza, Aly Fathy, and Jian Liu.
\newblock Mm-runassist: mmwave-based respiratory and running rhythm analysis during treadmill workouts.
\newblock {\em Proceedings of the ACM on Interactive, Mobile, Wearable and Ubiquitous Technologies}, 9(3):1--24, 2025.

\bibitem{gong2021rf}
Jian Gong, Xinyu Zhang, Kaixin Lin, Ju~Ren, Yaoxue Zhang, and Wenxun Qiu.
\newblock Rf vital sign sensing under free body movement.
\newblock {\em Proceedings of the ACM on Interactive, Mobile, Wearable and Ubiquitous Technologies}, 5(3):1--22, 2021.

\bibitem{asyari202460}
Rifa Atul~Izza Asyari, M~Zharfan Wiranata, Changzhi Li, Roy~BVB Simorangkir, and Daniel Teichmann.
\newblock 60 ghz fmcw millimeter wave radar assisted with dual-layer wideband flexible metasurface for accurate wrist pulse monitoring.
\newblock {\em IEEE Sensors Letters}, 8(3):1--4, 2024.

\bibitem{velloso2013qualitative}
Eduardo Velloso, Andreas Bulling, Hans Gellersen, Wallace Ugulino, and Hugo Fuks.
\newblock Qualitative activity recognition of weight lifting exercises.
\newblock In {\em Proceedings of the 4th Augmented Human International Conference}, pages 116--123, 2013.

\bibitem{lu2023investigation}
Yun Lu, Yudong Cao, Yi~Chen, Hui Li, Weihua Li, Haiping Du, Shiwu Zhang, and Shuaishuai Sun.
\newblock Investigation of a wearable piezoelectric-imu multi-modal sensing system for real-time muscle force estimation.
\newblock {\em Smart Materials and Structures}, 32(6):065013, 2023.

\bibitem{khurana2018gymcam}
Rushil Khurana, Karan Ahuja, Zac Yu, Jennifer Mankoff, Chris Harrison, and Mayank Goel.
\newblock Gymcam: Detecting, recognizing and tracking simultaneous exercises in unconstrained scenes.
\newblock {\em Proceedings of the ACM on Interactive, Mobile, Wearable and Ubiquitous Technologies}, 2(4):1--17, 2018.

\bibitem{wu2025bandei}
Hongrui Wu, Feier Long, Hongyu Mao, JaeYoung Moon, Junyi Zhu, and Yiyue Luo.
\newblock Bandei: A flexible electrical impedance sensing bandage for deep muscles and tendons.
\newblock In {\em Proceedings of the 38th Annual ACM Symposium on User Interface Software and Technology}, pages 1--12, 2025.

\bibitem{mahmoodi_echoforce_2025}
Kian Mahmoodi, Yudong Xie, Tan Gemicioglu, Chi-Jung Lee, Jiwan Kim, and Cheng Zhang.
\newblock {EchoForce}: Continuous grip force estimation from skin deformation using active acoustic sensing on a wristband.

\bibitem{mills2005basics}
Kerry~R Mills.
\newblock The basics of electromyography.
\newblock {\em Journal of Neurology, Neurosurgery \& Psychiatry}, 76(suppl 2):ii32--ii35, 2005.

\bibitem{de2006electromyography}
Carlo De~Luca.
\newblock Electromyography.
\newblock {\em Encyclopedia of medical devices and instrumentation}, 2006.

\bibitem{clancy2002sampling}
Edward~A Clancy, Evelyn~L Morin, and Roberto Merletti.
\newblock Sampling, noise-reduction and amplitude estimation issues in surface electromyography.
\newblock {\em Journal of electromyography and kinesiology}, 12(1):1--16, 2002.

\bibitem{kahl2015effects}
Lorenz Kahl, Marcus Eger, and Ulrich~G Hofmann.
\newblock Effects of sampling rate on automated fatigue recognition in surface emg signals.
\newblock {\em Current Directions in Biomedical Engineering}, 1(1):80--84, 2015.

\bibitem{durkin2005effects}
Jennifer~L Durkin and Jack~P Callaghan.
\newblock Effects of minimum sampling rate and signal reconstruction on surface electromyographic signals.
\newblock {\em Journal of Electromyography and Kinesiology}, 15(5):474--481, 2005.

\bibitem{li2010selection}
Guanglin Li, Yaonan Li, Zhiyong Zhang, Yanjuan Geng, and Rui Zhou.
\newblock Selection of sampling rate for emg pattern recognition based prosthesis control.
\newblock In {\em 2010 Annual International Conference of the IEEE Engineering in Medicine and Biology}, pages 5058--5061. IEEE, 2010.

\bibitem{silva2004mmg}
Jorge Silva, Winfried Heim, and Tom Chau.
\newblock Mmg-based classification of muscle activity for prosthesis control.
\newblock In {\em The 26th Annual International Conference of the IEEE Engineering in Medicine and Biology Society}, volume~1, pages 968--971. IEEE, 2004.

\bibitem{sharshar2023camera}
Ahmed Sharshar, Ahmed H~Abo Eitta, Ahmed Fayez, Mohamed~A Khamis, Ahmed~B Zaky, and Walid Gomaa.
\newblock Camera coach: activity recognition and assessment using thermal and rgb videos.
\newblock In {\em 2023 International Joint Conference on Neural Networks (IJCNN)}, pages 1--8. IEEE, 2023.

\bibitem{ganesh2020personalized}
Preetham Ganesh, Reza~Etemadi Idgahi, Chinmaya~Basavanahally Venkatesh, Ashwin~Ramesh Babu, and Maria Kyrarini.
\newblock Personalized system for human gym activity recognition using an rgb camera.
\newblock In {\em Proceedings of the 13th ACM International Conference on PErvasive Technologies Related to Assistive Environments}, pages 1--7, 2020.

\bibitem{zhang2011physical}
Hong Zhang, Lu~Li, Wenyan Jia, John~D Fernstrom, Robert~J Sclabassi, Zhi-Hong Mao, and Mingui Sun.
\newblock Physical activity recognition based on motion in images acquired by a wearable camera.
\newblock {\em Neurocomputing}, 74(12-13):2184--2192, 2011.

\bibitem{belbasis2015muscle}
Aaron Belbasis, Franz~Konstantin Fuss, and Jesper Sidhu.
\newblock Muscle activity analysis with a smart compression garment.
\newblock {\em Procedia Engineering}, 112:163--168, 2015.

\bibitem{wang2015smart}
Qi~Wang, Wei Chen, Annick~AA Timmermans, Christoforos Karachristos, Jean-Bernard Martens, and Panos Markopoulos.
\newblock Smart rehabilitation garment for posture monitoring.
\newblock In {\em 2015 37th Annual international conference of the IEEE engineering in medicine and biology society (EMBC)}, pages 5736--5739. IEEE, 2015.

\bibitem{alvarez2022towards}
Jonathan~T Alvarez, Lucas~F Gerez, Oluwaseun~A Araromi, Jessica~G Hunter, Dabin~K Choe, Christopher~J Payne, Robert~J Wood, and Conor~J Walsh.
\newblock Towards soft wearable strain sensors for muscle activity monitoring.
\newblock {\em IEEE Transactions on Neural Systems and Rehabilitation Engineering}, 30:2198--2206, 2022.

\bibitem{almohimeed2013development}
Ibrahim AlMohimeed, Hisham Turkistani, and Yuu Ono.
\newblock Development of wearable and flexible ultrasonic sensor for skeletal muscle monitoring.
\newblock In {\em 2013 IEEE International Ultrasonics Symposium (IUS)}, pages 1137--1140. IEEE, 2013.

\bibitem{zhicheng_mmWave_vitalsign}
Zhicheng Yang, Parth~H. Pathak, Yunze Zeng, Xixi Liran, and Prasant Mohapatra.
\newblock Monitoring vital signs using millimeter wave.
\newblock In {\em Proceedings of the 17th ACM International Symposium on Mobile Ad Hoc Networking and Computing}, MobiHoc '16, page 211–220, New York, NY, USA, 2016. Association for Computing Machinery.

\bibitem{gillani2023Tumor}
Nazia Gillani, Tughrul Arslan, and Gillian Mead.
\newblock An unobtrusive method for remote quantification of parkinson’s and essential tremor using mm-wave sensing.
\newblock {\em IEEE Sensors Journal}, 23(9):10118--10131, 2023.

\end{thebibliography}
\end{document}